\documentclass[iop,apjl,tighten]{emulateapj}

\newcommand{\mach}{\mathcal M}
\newcommand{\edot}{\dot{E}}
\newcommand{\pdot}{\dot{P}}
\newcommand{\chan}{{\sl Chandra}}
\newcommand{\xmm}{{\sl XMM-Newton}}

\def\gta{\ifmmode {\mathbin{\lower 3pt\hbox   %> or of order
    {$\,\rlap{\raise 5pt\hbox{$\char'076$}}\mathchar"7218\,$}}}
    \else {${\mathbin{\lower 3pt\hbox
    {$\rlap{\raise 5pt\hbox{$\char'076$}}\mathchar"7218\,$}}}
    $}\fi}
\def\lta{\ifmmode {\,\mathbin{\lower 3pt\hbox   %< or of order
    {$\,\rlap{\raise 5pt\hbox{$\char'074$}}\mathchar"7218\,$}}}
    \else {${\mathbin{\lower 3pt\hbox
    {$\rlap{\raise 5pt\hbox{$\char'074$}}\mathchar"7218\,$}}}
    $}\fi}

\shorttitle {New X-ray Observations of the Geminga PWN}
\shortauthors {Pavlov, Bhattacharyya, Zavlin}

\begin{document}

\title {New X-ray Observations of the Geminga Pulsar Wind Nebula}

\author {
George G. Pavlov\altaffilmark{1},
Sudip Bhattacharyya\altaffilmark{2}, and 
Vyacheslav E. Zavlin\altaffilmark{3}}

\altaffiltext{1}{Pennsylvania State University, 525 Davey Lab., University Park, PA
16802; pavlov@astro.psu.edu}

\altaffiltext{2}{Department of Astronomy and Astrophysics, 
Tata Institute of Fundamental Research, Mumbai 400005, India;
sudip@tifr.res.in}

\altaffiltext{3}{USRA/NASA MSFC Space Science Office, Huntsville, AL 35812;
vyacheslav.zavlin@msfc.nasa.gov}

\begin{abstract}
Previous observations of the 
middle-aged pulsar Geminga with {\sl XMM-Newton} and {\sl Chandra} 
have shown an unusual pulsar wind nebula (PWN), with a $20''$ long central 
(axial) tail directed opposite to 
the pulsar's proper motion
and two $2'$ long, bent lateral (outer) tails.
Here we report on a deeper
 {\sl Chandra} 
observation (78 ks exposure)
 and a few additional \xmm\ observations of the Geminga PWN.
The new {\sl Chandra} observation has shown that the axial tail,
which includes up to three brighter blobs, extends at least $50''$
(i.e., $0.06 d_{250}$ pc) 
from the pulsar ($d_{250}$ is the distance scaled to 250 pc). It also
allowed us to image the patchy
outer tails and the emission in the immediate vicinity of the
pulsar  with high resolution.
The PWN luminosity, $L_{\rm 0.3-8\, keV}\sim
3\times 10^{29}d_{250}^2$ erg s$^{-1}$,
is lower than the pulsar's
magnetospheric luminosity by a factor of 10.
The spectra of the PWN elements are rather hard
(photon index $\Gamma\sim 1$). 
Comparing the two {\sl Chandra} images, we found evidence of PWN variability,
including possible motion of the blobs along the axial tail.
The X-ray PWN is the synchrotron radiation from relativistic particles of
the pulsar wind; its morphology is connected with
the supersonic motion of Geminga.
We speculate that the  outer tails are either (1) a sky projection of
the limb-brightened boundary
of a shell
formed in the region of contact discontinuity, where the wind bulk flow
is decelerated by shear instability, or
(2) polar outflows from the pulsar bent by the ram pressure from the ISM.
In the former case, the axial tail may be a jet emanating along the
pulsar's spin axis, perhaps aligned with the direction of motion.
In the latter case, the axial tail may be
the shocked pulsar wind collimated by
the ram pressure.
\end{abstract}

\keywords{pulsars: individual (Geminga) --- stars: neutron --- stars: winds, outflows 
--- X-rays: stars}

\section {Introduction} \label{sec: 1}

Pulsars lose their spin energy via relativistic pulsar winds (PWs) of 
charged particles.
The PW shocks in the ambient medium and forms a pulsar wind nebula (PWN) whose
synchrotron radiation can be observed in a very broad energy range, from
the radio to TeV $\gamma$-rays
(see Kaspi et al.\ 2006, Gaensler \& Slane 2006, and Kargaltsev \& Pavlov 2008
[KP08 hereafter] for recent reviews).
The shocked PW is confined between the termination shock (TS) and contact
discontinuity (CD) surface that separates the shocked PW from the shocked
ambient medium between the CD and the forward shock (FS). 
The shapes of the TS, CD, and FS depend on the wind outflow geometry 
and the ratio of the pulsar's speed to the sound speed in the ambient
medium (the Mach number), $\mach=v_{\rm psr}/c_s$. In particular, if the pulsar
moves with a supersonic speed, $\mach\gg 1$, and the preshock PW is
isotropic, then the TS, CD, and FS acquire bow-like shapes ahead of the
pulsar, with the TS apex (``head'') at a distance 
$R_{\rm TS,h}\approx (\edot/4\pi c p_{\rm ram})^{1/2}$, where 
$p_{\rm ram}=\rho_{\rm amb} v_{\rm psr}^2$ is the ram pressure, $\rho_{\rm amb}$ the
density of the ambient medium (e.g., Bucciantini et al.\ 2005; hereafter B+05 ).
The shocked PW forms a 
tail behind the pulsar, with a flow speed
significantly exceeding the pulsar's speed 
(Romanova et al.\ 2005; B+05).

Among $\sim$60 PWNe detected
by \chan, about 20 PWNe show such bowshock-tail morphologies
(KP08). 
Such tails have been observed, for instance, behind the pulsars 
J1747--2958 (Gaensler et al.\ 2004), J1509--5850 (Kargaltsev et al.\ 2008),
B0355+54 (McGowan et al.\ 2006), and B1929+10 (Misanovic et al.\ 2008),
with very different spindown ages,  
$\tau_{\rm sd} \equiv P/(2\dot{P}) = 26$, 160, 620, and 3100 kyr, respectively. 
We should note, however, that the detailed shape of the detected bowshock-tail
PWNe is often different from the idealized models, especially in the immediate
vicinity of the pulsar, possibly because of anisotropies of the pulsar outflows.
For instance, by analogy with a few bright, well-resolved PWNe around
young pulsars moving with subsonic velocities (such as the Crab 
PWN; Weisskopf et al.\ 2000),
one can expect that the pulsar outflows consist of equatorial 
and axial components, with respect to the spin axis, which are responsible
for the ``tori'' and ``jets'' observed in these torus-jet PWNe (KP08).

One of the most peculiar PWNe has been detected around the 
famous Geminga pulsar
(PSR J0633+1746).
Geminga was discovered as a $\gamma$-ray source
$\gamma$195+5, with the
{\sl SAS-2} satellite
(e.g., Thompson et al.\ 1977).
The period of Geminga, $P = 237$ ms, was discovered by Halpern \& Holt (1992)
 in X-ray
observations  with the {\sl R\"{o}ntgen Satellit}
({\sl ROSAT}),
and the period derivative,
$\pdot = 1.1\times 10^{-14}$ s s$^{-1}$, was first measured by
Bertsch et al.\ (1992)
in $\gamma$-rays with the {\sl Compton Gamma Ray Observatory} ({\sl CGRO}).
The period and its derivative correspond to
the spindown age $\tau_{\rm sd} 
=340$ kyr
and spindown power
$\edot = 3.3\times10^{34}$ erg s$^{-1}$.
The Geminga pulsar has also been detected in the optical (Halpern \& Tytler 1988;
Bignami et al.\ 1988), near-IR (Koptsevich et al.\ 2001), and UV (Kargaltsev
et al.\ 2005). The distance to Geminga, $d=250^{+120}_{-62}$ pc, was
estimated from its annual parallax measured in observations with
the {\sl Hubble Space Telescope} (Faherty et al.\ 2007).
Its proper motion, $178.2\pm1.8$ mas/yr, corresponds to the transverse
velocity, $v_\perp \approx 211 d_{250}$ km s$^{-1}$ [where
$d_{250}=d/(250\, {\rm pc})$]. As this velocity
considerably exceeds the typical
sound speed in the interstellar medium (ISM), $c_s\sim 10$--30 km s$^{-1}$,
one should expect Geminga to be accompanied by a bowshock-tail PWN, with
$R_{\rm TS,h} =1.1\times 10^{16} n^{-1/2} (d_{250}/\sin i)^{-1}$ cm, which
corresponds to $\approx 2.9'' n^{-1/2} (d_{250}/\sin i)^{-2}$,
where $i$ is the angle between the pulsar's velocity and the line of sight,
and $n=\rho_{\rm amb}/m_{\rm H}$.

{\sl XMM-Newton} observations of Geminga in 2002 April,
reported by Caraveo et al.\ (2003; hereafter C+03),
revealed 
two $\approx 2^{\prime}$ long tails
behind the pulsar,
approximately symmetric with respect to the sky projection of the pulsar's
trajectory
(see Fig.\ 1), with a luminosity of $\sim 10^{29}$ erg s$^{-1}$ 
in the 0.3--5
keV band.
C+03 suggested that these tails 
are associated with a bowshock generated by the pulsar's motion, and,
using the one-zone bowshock model  by Wilkin (1996)\footnote{The one-zone 
model assumes
an instant mixing of the PW matter with the ambient matter at the shock,
so that there is no CD, and the TS and FS suraces coincide with each
other. The numerical bowshock models (e.g., Bucciantini 2002) have
shown that the shape of the one-zone shock is approximately similar to 
that of the FS but very different from the TS shape.},
predicted
that the head of the bowshock,
$20^{\prime\prime}$--$30^{\prime\prime}$
ahead of the pulsar,
is hidden in the bright wings of the pulsar point spread function 
(PSF) in the \xmm\ image.

The Geminga field was observed in 2004 (Sanwal et al.\ 2004; 
Pavlov et al.\ 2006 [hereafter P+06])
with the \chan\ Advanced CCD Imaging Spectrometer
(ACIS),
whose resolution,
$\approx 0\farcs5$, is much better than that of the \xmm\ detectors.
The most interesting finding of that observation was the  
detection of an axial tail behind
the pulsar aligned with the direction of the pulsar's proper motion
(P+06; de Luca et al.\ 2006; see Fig.\ 2, top).
The axial tail, with a luminosity $\sim 2\times 10^{29}d_{250}^2$ erg s$^{-1}$,
 was seen up to $25''$ from the pulsar, almost up to the
boundary of the field of view (FOV). 
P+06 suggested that the axial tail may be a jet emanating from the pulsar
magnetosphere.
In addition to the axial tail, a faint arc-like structure was
detected 
$5''$--$7''$ ahead of
the pulsar (but no emission at $20''$--$30''$, contrary to the C+03 prediction), and 
a $3\,\sigma$ enhancement, apparently connecting the
arc with one of the outer tails (south of the axial tail), 
was noticed (P+06). 
No emission was detected from the other (northern) outer tail in that
short, $\approx 20$ ks, exposure.

To image the whole extent of the Geminga PWN and study its tails in more
detail, we observed this field with \chan\ ACIS in 2007, with a longer exposure
and a larger FOV.
In this paper, we report the results of this observation
and compare them with the previous findings.
We describe the data analysis and the observational 
results in \S\,2, and discuss the implications of these results in \S\,3.

\vspace{1cm}

\section {Data Analysis and Results} \label{sec: 2}

\subsection{Observations and data reduction}

The Geminga field was observed with \chan\ ACIS on 2007 August 27
for 
78.12 ks (ObsID 7592). The observation was taken in Timed Exposure (TE)
mode, with the frame time of 3.24 s.
After removing 20 s of high background
and correcting for the detector dead time,
the scientific exposure time (live time) 
is 77,077 s.
To maximize the signal-to-noise ratio ($S/N$) for the very faint
PWN emission, we imaged the field onto the front-illuminated
 I3 chip, which has a 
lower background than the commonly used (and slightly more sensitive)
 back-illuminated
S3 chip\footnote{See \S\,6.16 of The Chandra Proposers' Observatory Guide (POG),
\url{http://asc.harvard.edu/proposer/POG}.}.
We used the Very Faint telemetry format to provide a better screening of background
events\footnote{See \S\,6.14 of the \chan\ POG.}.
As putting the target at the ACIS-I aimpoint (near the corner of the
I3 chip) could result in chip gaps crossing the PWN image, we moved
the focus to the middle of node 2 on the I3 chip
(SIM-Z $=-7.42\,{\rm mm} = - 2\farcm53$) and applied the
$\Delta Y=-1\farcm6$ offset to put the pulsar at $\gtrsim 2'$ from the chip
boundaries.

To obtain a deeper PWN image and examine a possible PWN variability, we also
used the previous \chan\ observation of Geminga carried out
on 2004 February 7 (ObsID 4674; 18,793 s scientific exposure ).
The details of that observation have been described by P+06. Here we only
mention that the observation was taken in Faint telemetry format, and the
Geminga pulsar and its PWN were imaged on 1/8 subarray of the S3 chip
($\approx 1'\times 8'$ FOV), which
reduced the pileup in the pulsar image but did not allow us to image the
whole PWN.

We have used the Chandra Interactive Analysis
of Observations (CIAO) software (ver.\ 4.0; CALDB ver.\ 3.4.0) for 
the ACIS data analysis, starting from the level 1 event files. We have applied
the standard grade filtering
and used the energy range 0.3--8 keV to minimize the background
contribution.
We have also applied
the exposure map correction, but found that the effects of nonuniform exposure and 
nonuniform CCD response in the PWN region are small (except for the boundary
of the 1/8 subarray used in the observation of 2004). 

To confront the high-resolution \chan\ data with the \xmm\ results, we also
used the data obtained with the MOS1  and MOS2 detectors
of the European Photon
Imaging Camera (EPIC) on board  \xmm. In addition to the observation of
2002 April 4--5 
(ObsID 011117010; 
77.97 ks scientific exposure, after removing
the periods of high background) reported by 
C+03 and P+06, we also
used the data sets obtained in observations 0201350101 of
 2004 March 13 (16.23 ks scientific exposure), 
031159100 of 2006 March 17 (4.46 ks), 
0400260201, 2006 October 2 (19.70 ks), and 
0400260301 of 2007 March 11 (23.83 ks). The total effective exposure
of the five observations is 142.18 ks.
All the observations were taken with medium filter in Full Frame mode, 
providing a $30'$ diameter FOV.
The data reduction was performed with the Scientific Analysis 
System (SAS) package (ver.\ 8.0.0). 
Good events with patterns 0--12 and energies
within the 0.3--8 keV range were selected for the data analysis.

\subsection{Images and spectra of the PWN elements}

The \chan\ ACIS data of 2007 provide the 
high-resolution image of the entire Geminga PWN 
for the first time (see the top panel of Fig.\ 2 and Fig.\ 3). 
Below we will describe 
the observed properties of the PWN elements, 
and compare them with the results of the 2004 ACIS
observation.

To calculate the net source counts $N_S$
in the area $A_S$ of a PWN element, we use the formula $N_S =
N_T - (A_S/A_B)N_B$, where $N_T$ is the total
number of counts detected from the area $A_S$, and $N_B$ is the number
of background counts detected from the area $A_B$. Then the
$1\, \sigma$ error of $N_S$ and the
signal-to-noise ratio are given by $\delta N_S =
[N_S + (1 + A_S/A_B)(A_S/A_B)N_B]^{1/2}$
and $S/N = N_S/\delta N_S$, respectively.

For the analysis of the 2004 data, we use a source-free rectangular background region
($A_B=
3439$ arcsec$^2$, $N_B=129$ counts in the 0.3--8 keV band)
to the north of the Geminga pulsar.
For the 2007 data, we 
use the background measured from 
a source-free rectangular region, with the area $A_B = 5399$ arcsec$^2$,
in the northeast portion of the ACIS-I3 chip (unless stated otherwise). 
This region contains $N_B=416$ counts 
in the 0.3--8 keV band, 
which corresponds to the background brightness of $1.0\times 10^{-6}$
counts arcsec$^{-2}$ s$^{-1}$, a factor of 2 lower than in the \chan\ observation of 2004.
The values of $N_S$ and $S/N$ for the PWN elements are given in Table 1.

For the spectral analysis, we have used
the XSPEC package (ver.\ 12.4.0) and fit the spectra with the
absorbed power-law (PL) model ({\tt wabs*powerlaw}), 
with the fixed hydrogen column density $N_{\rm H} = 
1.1\times10^{20}$ cm$^{-2}$ (Halpern \& Wang 1997; de Luca et al. 2005).
As the number of counts in the PWN is small, 
we use the maximum likelihood method (C-statistic) for spectral fitting.
Table 1 provides the values of the photon index $\Gamma$ and the flux
$F$ of the PWN elements.

\subsubsection{Axial tail}

The brightest feature of the Geminga PWN in the {\sl Chandra} data 
of 2004
is the axial tail (A-tail hereafter), 
seen up to at least $25''$ from the pulsar in the
direction opposite to the pulsar's proper motion
(see P+06 and Fig.\ 2, middle).
In the image from the 2007 observation, we see the 
A-tail up to at least $50''$ ($0.06\, d_{250}$ pc) from the pulsar
 (Fig.\ 2, top), with 
$83\pm 12$ source counts within the
 region of 
706 arcsec$^2$ area shown
by the solid lines in Figure 3. 
The PL fit of its spectrum 
(see Table 1 and Fig.\ 4) gives the photon index $\Gamma=1.8\pm 0.3$
and the 0.3--8 keV
luminosity $L= (0.9\pm 0.1)\times 10^{29}$ erg s$^{-1}$ 
(assuming an isotropic emission at $d=250$ pc),
versus $\Gamma=1.3\pm 0.3$ and $L=(1.6\pm 0.3)\times 10^{29}$ erg s$^{-1}$ 
in the 2004 data, 
as measured in the
118 arcsec$^2$ area rectangle that contains $46\pm 7$ counts
(shown in the middle panel of Fig.\ 2).

The A-tail looks patchy in both the 2004 and 2007
observations, with some ``blobs'' standing out.
The blobs, labeled A, B, and C in Figure 2
are at the 
distances of about $9''$, $20''$, and $43''$
from the pulsar, 
respectively 
(blob A is seen in the 2004 image, while blob B and blob C are seen
in the 2007 image).
They contain
$24.9\pm 5.1$, $12.8\pm 3.9$, and $25.8\pm 5.3$ source counts, respectively,
within the $3''$-radius circles around their centers.
The analysis of the brightness distribution along the A-tail shows that 
the blobs are significant at $>3\sigma$ levels (i.e., they
are not just statistical fluctuations of the brightness distributions).
For instance, the number of
counts in the $2''$ radius circle around the center of blob B
(15 counts)  exceeds
the average number of counts per the same 12.6 arcsec$^2$ area
in the A-tail ($2.44\pm 0.21$ counts) at the $3.2\,\sigma$ level.
The nonuniform surface brightness distribution along the A-tail,
and the difference of these distributions in the 2004 and 2007 images
are 
shown in Figure 5.

The brightest in the 2007 data is blob C at the apparent end of the tail,
centered at
$\alpha=06^{\rm h}33^{\rm m}51\fs71$, $\delta= +17^\circ 45' 53\farcs1$ (J2000).
Because of the small number of counts, we cannot firmly determine whether
the blob corresponds to a point source or an extended one.
Interestingly, the end portion of the tail looks attached
to this blob, while the tail looks detached from the pulsar in both
the 2004 and 2007 images. Therefore, one could even speculate that the tail might belong
not to Geminga but to some unrelated field object (e.g., it might be a jet
of an active galactic nucleus [AGN], accidentally oriented toward Geminga
in the sky projection).

To check whether the blobs are indeed associated with the tail or they may be
 background sources, we examined the optical/NIR catalogs.
We found no optical counterparts to blob A and blob B, but we 
found  
an object at $\alpha=06^{\rm h}33^{\rm m}51\fs69$,
$\delta= +17^\circ 45' 54\farcs2$ (J2000), 
about 
$1''$ from the center of blob C
(the coordinates are from the USNO-B1.0 catalog
[Monet et al.\ 2003], with the
quoted 
mean uncertainties of $0\farcs075$ and $0\farcs094$ in $\alpha$ and
$\delta$, respectively).
Based on the magnitudes and colors (e.g., $V=17.7\pm 0.3$
[GSC2.3 catalog; Lasker et al.\ 2008], 
$J=16.18\pm 0.09$, $H=15.54\pm 0.11$, $K=15.6\pm 0.2$
[2MASS catalog; Cutri et al.\ 2003]), this object could be a background 
K star. Such a star could contribute to the X-ray emission of blob C.
The observed X-ray flux in the $3''$ radius aperture 
 is $F_{\rm 0.3-3.5\,keV}\approx
3.8\times 10^{-15}$ erg cm$^{-2}$ s$^{-1}$. 
According to Maccacaro et al.\ (1988), the X-ray/optical flux ratio,
 $\log (F_{\rm 0.3-3.5\,keV}/F_V)
\approx -2.0$, corresponds to a K or M star, and it excludes an AGN
[for which $-1.2 < \log (F_{\rm 0.3-3.5\,keV}/F_V) < +1.6$)
as the source of the X-ray and optical emission
(hence the tail is not an AGN jet).
Thus, 
we cannot rule out the possibility that a K star, accidentally projected onto 
the A-tail, is at least partly responsible for the brightened end of the
tail in the 2007 image.
However, if 
we exclude blob C,
the tail's luminosity
decreases by a factor of 1.5, but the spectral slope remains virtually
the same (see Table 1 and Fig.\ 4). This suggests that the
star's contribution does not dominate in the 
blob C emission, but, because of the small number of counts and large statistical
errors, we cannot firmly conclude on the nature of blob C.

In the observation of 2004, 
blob C was imaged onto an underexposed
part of the FOV (because of the dither), 
between the dashed lines in the middle panel of Figure 2.
 Taking into
account the shorter effective exposure of that observation
 (but the higher sensitivity
of the S3 chip), we expect $4.2\pm 2.1$ counts to be detected
in the $2''$ radius circle around the position of
the blob C centroid; 
however, 
there are no counts within that circle.
This may suggest some variability of the source,
but the statistical significance of this difference
is marginal (e.g., the probability of detecting zero counts when 4.2 counts
are expected is 0.0145, which corresponds to a $2.4\, \sigma$ significance).

Based on the ACIS count rate of blob C in 2007,
one could expect to detect about 90 counts in the $15''$
radius aperture in the 142 ks MOS1+MOS2 exposure, but we found only
$49\pm12$ counts in the MOS data.
Furthermore,
we note that the blob C position is projected onto the 
wings of the pulsar PSF in the \xmm\ images (see Fig.\ 1),
whose contribution to the number of extracted counts
is difficult to evaluate because of the ``spiky'' shape of the PSF.
 Anyway, the number 
of counts expected for blob C in the \xmm\ data
significantly exceeds 
the measured one, suggesting
variability of blob C.

The A-tail images (Fig.\ 2) 
look appreciably different in the 2004 and 2007 data, in both
the overall flux and the surface brightness distribution (see Fig.\ 5).
For instance, the flux, $2.2\times 10^{-14}$ erg cm$^{-2}$ s$^{-1}$,
in the 118 arcsec$^{2}$ area of the tail in the 2004 data is a factor of
 6 higher than the flux from the same area
in the 2007 data 
(the difference between the count rates 
is significant at the $5\,\sigma$ level, with
account for a factor of 1.6 higher sensitivity of the S3 chip
compared to that of the I3 chip, for $\Gamma=1.8$ and $N_{\rm H}=1.1\times 10^{20}$ cm$^{-2}$).

The different positions of the blobs in the 2004 and 2007 images suggest
that the blobs are moving along the A-tail (perhaps similar
to the blobs in the Vela pulsar jet; Pavlov et al.\ 2003).
One might even speculate that, for instance, blob B is, in fact, blob A that had
 moved $11''$ ($4\times 10^{16}\, d_{250}$ cm) in 3.5 yr
between the observations
(which would correspond to the transverse
velocity of $v_{\rm blob, \perp}\sim 3700\, d_{250}$ km s$^{-1}$).
However, as blob B could also form independently after the disappearance of
blob A, this will remain a speculation until the 
characteristic blob lifetime is estimated in a series of monitoring
observations.

In the large-scale {\sl Chandra} image of 2007 (Fig.\ 3),
one can see
a possible faint
extension of the A-tail
(within the dashed polygon in Fig.\ 3).
This faint portion has a factor of 1.5 higher
observed flux (but a factor of 2.4 lower average brightness)
than the bright portion (see Table 1). 
However, its statistical significance is only $1.7\,\sigma$, and it
is not seen
in the deep \xmm\ image (see Fig.\ 1). Therefore,  
this ``faint portion'' most likely represents a string of background
fluctuations accidentally aligned in the A-tail direction.
 
\subsubsection{Outer tails}

The two ``outer tails'' of the Geminga PWN, seen up to $\sim 2'$ from
the pulsar, were originally detected in the
\xmm\ observations of 2002 April (C+03). Adding four shorter 
observations of 
2004--2007, which increases the total exposure by a factor of 1.8, shows
qualitatively the same picture (Fig.\ 1). The two tails are approximately
symmetric with respect to the pulsar's trajectory in the sky, forming
a horseshoe-like structure. The southern and northern tails (we will call
them the S-tail and N-tail, for brevity) are seen up to 
$3\farcm1$ and $2\farcm7$ from the pulsar,
respectively, in the summed MOS1+MOS2 image.
Their typical width, $\sim 20''$--$30''$, is comparable to the \xmm\
angular resolution. The tails in the pulsar vicinity (within 
$\approx30''$) are 
immersed in the bright pulsar's 
image.
The spectrum of the combined emission from the two tails,
 extracted from the
$120''\times 40''$ elliptical regions
shown in Figure 1 (about 560 source counts),
can be described by a PL model with  $\Gamma=1.7\pm0.2$
($\chi_\nu^2 = 1.36$ for 50 degrees of freedom [d.o.f], for fixed $N_{\rm H}
=1.1\times 10^{20}$ cm$^{-2}$).
The unabsorbed flux and
luminosity of the two tails, 
 are
$F^{\rm unabs}\simeq 1.7\times 10^{-14}$ erg cm$^{-2}$ s$^{-1}$ and
$L\simeq 1.3\times 10^{29} d_{250}^2$ erg s$^{-1}$, in the 0.3--8 keV band.
Both the total flux and the spectral slope are
consistent with those obtained by C+03.
The contribution of the S-tail into the total energy flux is about 76\%.
The average specific
intensities in the S-tail and N-tail elliptical regions
 are $3.4\times 10^{-18}$ and
  $1.1\times 10^{-18}$ 
erg cm$^{-2}$ s$^{-1}$ arcsec$^{-2}$, respectively.
Fitting the spectra of the S-tail and N-tail separately, we obtained
$\Gamma=1.4\pm0.3$ and $2.7\pm 1.1$, 
respectively. The apparently large difference
between the spectral slopes, $\Delta\Gamma = 1.3\pm 1.1$, is not
statistically significant.
The spectra can also be fitted by the models for emission of an optically thin
thermal plasma (e.g, $kT=8^{+10}_{-3}$ keV for the fit
of the S-tail + N-tail spectrum with the {\tt mekal}
model; $\chi_\nu^2=1.36$ for 50 d.o.f.).
 
Thanks to its high angular resolution,
\chan\ observations 
 make it possible to image the tails in the pulsar vicinity 
and resolve the tail structure. In the short
 observation of 2004 the N-tail
was not detected, while the initial portion of the S-tail was detected
with about $3\,\sigma$ significance (P+06). The entire extent of the tails 
could not be seen because the 1/8 subarray was used.

In the \chan\ data of 2007 we have detected 
both outer tails.
For the analysis, we divide each of the tails into two parts.
The bright initial parts (up to $2'$ and $40''$ from the pulsar,
 for the S-tail and N-tail, respectively) are delineated by solid polygons
in Figure 3, while the longer faint parts
(up to  4\farcm2 and 3\farcm6 from the pulsar, respectively) 
are shown by dashed
polygons. 
(We note that the bright portion of the N-tail and a substantial part
of the bright portion of the S-tail are hidden behind the 
pulsar image in the \xmm\ data.)
The statistical significance of the faint parts is marginal in
the \chan\ data (see Table 1), but 
their reality 
is supported by the \xmm\ data (see Fig.\ 6,
where the \xmm\ brightness contours are overlaid on the \chan\ image).
The end parts of the S-tail in the \chan\ and \xmm\ images
are slightly shifted with respect to each other, which might suggest
possible variability of the outer tails.
The \chan\ image resolves the tails in the transverse dimension,
showing $\gtrsim 20''$ widths, but
the image is not deep enough to infer the brightness distribution across
the tails. Also, the tails do not show sharp-cut outer boundaries,
perhaps because of the same reason.

Figure 6 also shows the locations of four optical-NIR
sources that are projected onto the tails.
We have already discussed one of them (\#1 in Fig.\ 6,
 whose position coincides
with blob C in the 
A-tail).
The other three sources (numbered 2, 3, and 4 in Fig.\ 6,
with V magnitudes of 17.6, 15.3, and 13.4, respectively,
from the NOMAD catalog; Zacharias et al.\ 2004)
are  projected onto the S-tail. 
Source 3 and source 4 are likely an F star and a K star,
respectively, based on their
optical-NIR colors, while the magnitude errors for the fainter source 2 
are too large to determine its nature.
%magnitudes), while the other one, which is relatively
The $2.55''$ radius circles around the positions of the sources 2, 3, and 4
contain 7, 1, and 11 counts, respectively.
Even if the X-ray emission at these locations is due to
the optical sources, the
 total number of detected source counts from all the
three sources (assuming they are pointlike)
is only $14.3\pm4.4$,
 while
the number of counts from the entire S-tail is $168\pm 34$ (Table 1). Therefore,
their contribution to the observed X-ray emission from the entire S-tail is
negligible. 

The PL fits of the outer tails' spectra
(see Table 1 and Fig.\ 4) indicate
that the spectral slope of the S-tail  
does not differ significantly from that of the A-tail (e.g., $\Gamma \approx 1.5$ 
fits both spectra within the $1\, \sigma$ uncertainties), 
while  the N-tail is 
apparently harder, in contradiction to the result found from the \xmm\ data.
Although the observed number of counts from the N-tail is 
a factor of 1.8 lower than that
from the S-tail, 
their luminosities are comparable.
Table 1 also suggests that both the N-tail and S-tail spectra soften from the
bright parts toward the extended faint portions, but the statistical 
significance of the softening is low (e.g., $\Delta\Gamma\approx 0.7\pm 0.5$ for
the S-tail).
The total 
luminosity of the two tails is
$L\approx 4.2\times 10^{29}\,d_{250}^2$ erg s$^{-1}$,
in the 0.3--8 keV band, assuming isotropic emission. 
This value exceeds 
the estimate derived above from the
\xmm\ data
by a  factor of about 3,
but that estimate was obtained for a fraction of outer tails,
which did not include the bright part of the N-tail and included
only a small portion of the bright part of the S-tail.
The average specific 
intensities in the bright parts are about
$1.2\times 10^{-17}$ and $2.6\times 10^{-17}$ erg
cm$^{-2}$ s$^{-1}$ arcsec$^{-2}$, for the S-tail and N-tail, respectively.

Similar to the A-tail,
the bright and faint portions of the S-tail 
and the faint portion of the N-tail look patchy. 
To quantify the statistical significance of the patchiness, we have
compared the net counts from 
brighter and fainter regions of equal size for each of these
components. 
For the bright S-tail, faint S-tail, and faint N-tail,
we found significancies of $2.5\,\sigma$, $2.2\,\sigma$, and $2.1\,\sigma$,
respectively.
Therefore, the patchiness is not ruled out, but a deeper observation
is required to prove it firmly.  
A similar analysis of the \xmm\ data does not show a statistically
significant patchiness, 
because of large noise and poor angular resolution.

To examine variability of the outer tails, we have compared the count
rates in the area of 
462 arcsec$^2$ of the bright part of the N-tail ($39.4\pm 8.8$ and
$5.7\pm 5.0$ net counts in the observations of 2007 and 2004, respectively).
Accounting for the factor of 1.2 higher sensitivity of the S3 chip
compared to the I3 chip (for the spectral parameters derived from the 2007
observation), the difference between the count rates is significant
at the $3.5\,\sigma$ level.
On the other hand, the difference of the count rates in the
985 arcsec$^2$ area of the bright part of the S-tail,
which was detected in both the 2004 and 2007 \chan\ observations
($33.0\pm9.0$ and $79.1\pm13.0$ net counts, respectively),
is statistically insignificant.
We have also looked for variability of the outer tails in the \xmm\ data,
but found no statistically significant differences between the separate
observations because of the strong noise.

As it is natural to assume that the ``outer tails'' represent
the sky projection of
limb-brightened boundaries of a shell,
one can expect some X-ray emission from the region
between the outer tails, in addition to the A-tail. 
To look for this emission, we have 
inspected two inter-tail regions of combined area 
$A_S=12,898$ arcsec$^2$ that exclude
the entire (bright plus faint) A-tail and contain $N_T=886$ counts. 
Using the background extracted from three source-free
rectangles around the PWN ($N_B=2222$ counts in the combined area 
$A_B=29,578$ arcsec$^2$),
we found $N_S=-88\pm36$ 
 net counts from the ``source". 
Adding the alleged faint portion of the
A-tail,
which gives $N_T=1101$ counts in the area $A_S=15,302$ arcsec$^2$,
we found $N_S=-48\pm41$ net counts.
Thus, we conclude that there is no detectable
emission from the region between the outer tails, and the ``faint portion''
of the A-tail is likely an illusion (in agreement with our conclusion
at the end of \S\,2.2.1). 
Using the approach outlined by Weisskopf et al.\ (2007), we find the 
$3\,\sigma$ upper limit of 
94 counts in the area $A_S=15,302$ arcsec$^2$
(99\% and 90\% upper limits are 
78 and 45 respectively).
The corresponding $3\,\sigma$ upper limit on 
the surface brightness, $6.1\times 10^{-3}$ counts arcsec$^{-2}$,
 is lower than the average surface brightness of the
outer tails 
(e.g., by  factors of $4.6\pm 0.9$ and $3.4\pm 1.0$ for the
``entire''   S-tail
and N-tail, respectively).

The \xmm\ image also does not show detectable inter-tail emission.
Using the same approach, we found the $3\,\sigma$ upper limit of 
37 counts
in the $32''\times 64''$ box between the outer tails shown in Figure 1
(99\% and 90\% upper limits are 
32 and 22 counts, respectively).
The corresponding $3\,\sigma$ upper limit on surface brightness, $1.8\times 10^{-2}$ counts arcsec$^{-2}$,
 is lower by factors of $6.2\pm 0.2$ and $2.0\pm 0.4$
  than the surface brightnesses
of the S-tail and N-tail, respectively.

\subsubsection{Emission ahead of the pulsar}
An arc-like diffuse emission region, about $5''$--$7''$ ahead of the pulsar, 
was reported by P+06 from the 2004 {\it Chandra} data. 
We have analyzed the data inside a polygon (area $= 86.6$ arcsec$^2$) 
and found $22.8\pm5.1$ 
source counts.
Our spectral analysis provides a photon index
$\Gamma=0.9\pm0.5$ and the observed flux 
$F\approx 1.2\times10^{-14}$ erg cm$^{-2}$ s$^{-1}$,
in the 0.3--8 keV band, consistent with the P+06 estimates.

Based on these results, 
the expected number of arc counts in the \chan\ observation of 
2007\footnote{Estimated with the aid of the \chan\ PIMMS tool; 
\url{http://asc.harvard.edu/toolkit/pimms.jsp}}  is $83\pm 24$.
However, although some diffuse emission is seen at that site in the 2007  
data, there are only $16.3\pm4.8$ source counts in the corresponding polygon
(at the same distance from the pulsar, which has moved $0\farcs62$ in the sky in the
3.5 years),
and the shape of the count distribution does not resemble an arc
(see the upper panel of Fig.\ 2). 
The spectral slope, 
$\Gamma=1.1\pm0.7$, is  
consistent with that of the 2004 arc, but the observed flux, 
$F\approx 2.5 \times10^{-15}$ erg cm$^{-2}$ s$^{-1}$, 
is a factor of 5 lower.
This suggests that the emission ahead of the pulsar is
variable, but the significance of this variability is not very high
(e.g., $2.8\,\sigma$ in the difference between the expected and observed
counts).

The alleged arc cannot be seen in the \xmm\ images because it is hidden 
in the pulsar PSF. However, these images show a ``streak'' ahead of the pulsar,
in the direction of the proper motion, which is best seen in the summed image
(bottom panels of Fig.\ 1). One might speculate that this streak is 
a Geminga PWN element
 (e.g., a forward jet). To check this hypothesis, we extracted
760 source counts from the $20''\times 50''$ rectangle that includes the
streak (shown in the bottom panels of Fig.\ 1) and fit the spectrum with various
models. We found that the very soft streak spectrum does not fit the PL model
($\chi_\nu^2=4.3$ for 44 d.o.f.), but it fits the two-component 
PL+blackbody model ($\Gamma=1.8\pm 0.1$, $kT=0.53\pm 0.05$ keV,
$\chi_\nu^2=1.1$ for 42 d.o.f.) that is
consistent with the pulsar's spectrum (e.g., Kargaltsev et al.\ 2005).
Therefore, we conclude that the streak is not related to the PWN, but it
is an artificial spike-like feature 
in the MOS PSF caused by the ``spiders'' holding
the X-ray telescopes (see the XMM-Newton Users' Handbook, Sec.\ 3.2.1.1)
This conclusion is also supported by a lack of any excess above the
background at the corresponding area in the \chan\ images.

Interestingly, the summed 2004 + 2007 \chan\ 
image in the bottom panel of Figure 2
shows a hint of a short, $10''$--$15''$, jetlike structure ahead of the
pulsar, with its end seemingly connecting to the N-tail.
The number of
counts is, however, too small to conclude whether this structure is
an accidentally aligned superposition of events from the two
images or there is indeed a forward jet, perhaps bent in the north-northwest
direction into the N-tail, which we cannot see in the separate images 
because of the scarce statistics. A deeper observation is needed to 
understand the true nature of this and other apparent structures
in the immediate vicinity of the pulsar. 

\section {Discussion} \label{sec: 3}

The \chan\ observation of 2007 has confirmed the
existence of three tail-like features
in the
Geminga PWN image, with comparable luminosities,
  and allowed us to study the PWN in more detail
(e.g., to detect the A-tail at larger distances from the pulsar
and investigate the PWN in the immediate vicinity of the pulsar).
Moreover, it has provided first evidence of variability of the PWN elements,
in particular, of the A-tail and the emission in the pulsar vicinity.

The observed structure of the Geminga PWN looks very unusual.
Although the overall appearance of the PWN,
particularly the alignment of the tails with the pulsar's proper
motion, leaves no doubts that the PWN structure is caused by the 
supersonic motion of the pulsar in the ISM,
none of the other $\sim$20 bowshock-tail PWNe detected by \chan\ (see KP08)
show three distinct tails. 
To interpret the observed structure in terms of the PWN models,
 we should first understand the intrinsic
three-dimensional morphology of the X-ray PWN, which is by no means 
obvious. 

At the first glance,
the most natural interpretation of the PWN elements is that the ``outer
tails'', together with the ``arc'' that apparently connects the tails 
ahead of the pulsar,  delineate the limb-brightened boundary
of the sky projection
of an optically thin 
shell, shaped 
approximately as a paraboloid of revolution, while the nearly straight
axial tail represents 
a collimated outflow
in the direction opposite to that of the pulsar's motion. On the
other hand, 
one cannot  exclude the possibility
that the outer 
tails are,
in fact, hose-like
structures, such as jets confined by their own magnetic fields and bent
by the head wind of ISM matter.
Moreover, 
one could even speculate that the axial ``tail'' is Doppler-boosted
emission from a narrow region of a shell
formed by material flowing with relativistic speeds.
Below we will discuss
these possibilities in more detail.

\subsection{Possible interpretations of the outer tails}

The patchy outer tails of $20''$--$30''$ width are seen up to $\sim 3'$
($\sim 0.2 d_{250}$ pc) from the pulsar. Their X-ray luminosity,
$L_X\sim 4\times 10^{29} d_{250}^2$ erg s$^{-1}$ in the 0.3--8 keV  band,
is a fraction of $\sim 1.2\times 10^{-5}d_{250}^2$ of the pulsar's 
spin-down power $\edot$, lower than the typical $L_X/\edot \sim 10^{-4}$--$10^{-3}$ for younger PWNe (KP08). The outer tails' luminosity is a factor of       
$\sim 7$ lower than the pulsar's magnetospheric luminosity, while
the PWN luminosity is usually higher than the magnetospheric
luminosity for younger pulsars (Kargaltsev et al.\ 2007). The
spectrum of the outer tails can be described by the absorbed PL model with
$\Gamma \sim 1$, which is apparently harder than the typical PWN spectra.
The explanation of the outer tails' properties depends on the topology of
the PW outflow.

\subsubsection{The outer tails are shell boundaries?} 

Let us assume that the outer tails are 
limb-brightened shell boundaries and explore
the consequences of this assumption.
First of all, the very fact that the shell boundaries are much brighter
than the rest of the shell image
(see \S\,2.2.2) implies a nonrelativistic speed
of the bulk outflow along the shell.
If it were relativistic,
then, due to the Doppler boosting, the brightest parts of the
shell image would be not the boundaries
but they would correspond to the smallest angles between the flow velocity
and the line of sight.
This inference restricts the number of possible
interpretations of the shell.
For instance, the shell cannot be interpreted as
synchrotron emission from the shocked PW immediately outside
the (bullet-like) TS surface -- not only because the cylindrical radius of the
shell is too large (see P+06), but also because the shocked PW
 is expected to flow
 with nearly relativistic speed, $v_{\rm flow} =0.8c$--$0.9 c$,
along the lateral TS boundary (B+05). As the brightness is proportional to
$[1-(v_{\rm flow}/c)\cos\theta]^{-\Gamma-2}$, where $\theta$ is the angle
between the flow direction and the line of sight, the flow toward the
observer would be a factor of $\gtrsim 100$ brighter than the flow in the
perpendicular direction, for the photon index $\Gamma\approx 1$.

One might consider the possibility that the shell
emission comes from
the shocked ISM material heated up to X-ray temperatures.
In this case,
we would associate the outer tails with the FS, and the emission mechanism
with the thermal emission of an optically thin plasma rather than synchrotron
emission from relativistic electrons. From the junction conditions at
the shock front, the expected temperature of the shocked ISM material at
the head of the bowshock is
$kT = (3/16)\mu m_p V_{\rm psr}^2 =
0.078 \mu V_{200}^2$ keV
[where $\mu$ is the chemical weight, 
$V_{200}=V_{\rm psr}/(200\,{\rm km\,s}^{-1})$, and adiabatic index
$\gamma_{\rm ad} =5/3$ is assumed for the ISM gas],
and it should be even lower behind the pulsar because of the
obliqueness of the shock.
Since the fit of the outer tails spectrum with the {\tt mekal} model
gives $kT\approx 5$--18 keV (see \S\,2.2.2),  
the expected temperature of the shocked ISM gas looks too low to explain the hard spectrum of the
tails emission
unless $V_{\rm psr}\gtrsim 2300 \mu^{-1/2}(kT/10\,{\rm keV})^{1/2}$ km s$^{-1}$, which
would imply that the pulsar
moves at a small inclination angle $i$ with respect to the line of sight,
$\sin i \lesssim 0.1$.
However, since such a speed is higher than those observed for other pulsars,
and
 the PWN appearance can hardly be reconciled with such small inclination angles,
we can discard this interpretation.

It seems more reasonable to assume that the shell 
is formed by the shocked PW
 flowing immediately inside the CD surface, where the magnetic field is
compressed (B+05)
and the synchrotron radiation is enhanced.
However, such an interpretation contradicts the available
PWN models, which
predict a nearly relativistic
flow speed in the outer layers of the synchrotron
emitting PWN
(hence dim boundaries and bright central part)
because of the
Doppler boosting (see above). 
Indeed, the simulated PWN images for $i=90^\circ$ (see Fig.\ 4 
in B+05) show the brightest synchrotron emission from the
bowshock head region, while throughout the
entire PWN the brightness decreases from the axis 
toward the CD, in contrast to the observed images.
(This is partly caused by the assumption that the PWN magnetic field
is purely toroidal, which reduces the synchrotron intensity at the PWN
boundaries, where the magnetic field 
is parallel to the line of sight.
However, even if the magnetic field is
completely disordered, the brightness does not grow from the axis 
toward the boundaries; see Fig.\ 5 in B+05.)
Therefore, this interpretation of the outer tails implies
that there is a mechanism that decelerates the flow.
The deceleration can be provided by
the shear (Kelvin-Helmholtz) instabilities at the CD, which 
can lead to advection of clumps of the heavier shocked ISM material into the
shocked PW and slow down the latter (e.g., B+05). 
The diffuse appearance and the patchiness
 of the tails in the high-resolution \chan\
images (see \S\,2.2.2 and Figs.\ 2 and 3) are consistent with 
this hypothesis. 
It could be verified observationally
 if the speed of the nonuniformities in
the outer tails, which should form in the process of mass loading, is
measured in a series of deep observations. 

Another apparent problem with the interpretation of the outer tails
and the arc as traces of the CD surface
  is the discrepancy between the observed
and predicted ratios of the CD's cylindrcal radius,
$r_{\rm CD}$, to
the distance $R_{\rm CD,h}\approx 1.3 R_{\rm TS,h}$ of the 
CD head from the pulsar.
For instance,
the B+05 model predicts $r_{\rm CD}/R_{\rm CD,h}\approx 3$ while the observed
ratio is $\gtrsim 10 \sin i$ (if we interpret the 
``arc'' ahead of the pulsar as the head
of the CD surface\footnote{ If the PW is anisotropic, 
the head of the CD surface could
be much closer to the pulsar than the observed arclike emission. In this case,
the emission ahead of the pulsar could be a forward axial outflow destroyed
by the ISM ram pressure. Such an outflow should be seen as a diffuse 
emission of an irregular, variable shape, in contrast to the CD surface
head that would preserve its arclike shape and could show variations
 if the ISM is nonuniform.}.)
This means that either the inclination angle is small, 
$i < 15^\circ$--$20^\circ$,
or  some assumptions of the B+05 model are violated.
Since the observed PWN shape
can hardly be reconciled with such small inclination
angles (see Fig.\ 3 in C+03), we suggest that the discrepancy
is caused by the assumption that the unshocked PW is isotropic.
Indeed, if the wind is predominantly equatorial (i.e., concentrated
around the plane perpendicular to the pulsar's spin axis),
and the spin axis is aligned with the direction of pulsar motion
(as observed in a number of young pulsars), then 
the lower wind ram pressure ahead of the pulsar should result in a smaller
distance between the pulsar and the TS (and CD) apex.
The only models of PWNe with anisotropic wind outflow we are aware of
have been presented by Vigelius et al.\ (2007).
These models consider only nonrelativistic flows and, more
importantly, assume zero magnetic field, but they 
should satisfactorily describe
the PWN morphology for small values of the PW magnetization
parameter (defined as the ratio of the Poynting flux to the kinetic energy flux).
Although Vigelius et al.\ did not directly consider the case of
an equatorial outflow perpendicular to the pulsar velocity, some of the
considered cases (e.g., Fig.\ 6 in that paper) 
qualitatively confirm our explanation.

For completeness, we should also mention the interpretation suggested by C+03,
that the outer tails represent the synchrotron radiation of the PW
in the interstellar magnetic field ``compressed in the bowshock'' by
a factor of 4 (for an adiabatic shock with $\gamma_{\rm ad}=5/3$ and a large
Mach number), up to $\sim 10\,\mu$G. As C+03 assume
a one-zone shock model,
which apparently
does not describe realistic PWNe, this interpretation may not be directly
applicable. One may speculate, however,
 that ultrarelativistic electrons from
the high-energy tail of the electron energy distribution could leak
 from the shocked wind region (incide the CD) into the shocked ISM region 
and generate synchrotron radiation in the interstellar
magnetic field amplified at the FS.
We should note, however, that only the magnetic field component
parallel to the shock surface is amplified by this mechanism, and the
amplification becomes insignificant behind the pulsar because of
the shock obliqueness. Therefore, we believe that  there is no need 
to invoke this complicated
hypothesis as long as the more straightforward explanation
(synchrotron
radiation from the region of the CD surface) seems viable.

Finally, we should explain the fact that the outer tails are not truly
symmetric with respect to the trajectory of Geminga on the sky, 
neither in shape (especially close to the pulsar) nor, particularly,
 in brightness (the
S-tail is considerably brighter). We could tentatively ascribe this asymmetry
to nonuniform conditions (density and/or temperature)
 in the ambient medium. 
The nonuniformity is supported by the 
Very Large Array (VLA) and Effelsberg radio telescope
HI (21 cm line) observations of the Geminga field
(Giacani et al.\ 2005). These observations
have shown the pulsar and its X-ray PWN 
to be in a local minimum of the HI emission,
surrounded by an open HI shell (an incomplete ring with an average radius
of $9'$) that envelopes the southern part of the X-ray PWN, with the internal
border of the shell close to the S-tail. The lack of neutral hydrogen in the
vicinity of the pulsar can be explained by the ionization caused by
the pulsar's UV and
soft X-ray emission. The openness of the shell (no HI emission northwest
of the X-ray PWN) might imply a higher temperature (and perhaps a lower density)
of the ISM in that direction. One may speculate that Geminga is 
crossing
a cold ISM cloud and is now
approaching the cloud's northwest boundary. It remains to 
be understood, however, how the relative brightness of the S-tail is connected
with the alleged lower temperature and higher density in that region.
We should also mention that, based on the HI radio results, one could expect an
H$_\alpha$ PWN south of the S-tail, associated with the FS. However, C+03
report the nondetection of ``organized diffuse H$_\alpha$ emission from the
X-ray structure surrounding Geminga'' in a 5 hour observation with
the VLT-ANTU telescope. Their Figure 2 shows an apparent filament
at the outer border of the S-tail, but it is not immediately 
clear whether or
not this feature is related to the PWN. 

To conclude, if the outer tails represent the limb-brightened boundaries
of the sky projection of a shell, this shell is most likely the synchrotron
radiation from the region of interaction of the shocked PW and
shocked ISM material, where the wind flow is decelerated to nonrelativistic
velocities by the shear instability, which implies mass loading. 
The shape of the shell is somewhat
different from the shape of the CD surface in the available 
numerical PWN models, perhaps
because the models do not include the mass loading and proper anisotropy
of the unshocked PW.

\subsubsection{The outer tails are bent polar outflows?}

The hypothesis that the outer ``tails'' represent the boundary of
a shell 
is not the only possible explanation.
In particular, as no emission is seen between the outer tails 
except for the A-tail (see \S2.2.2),
we cannot exclude the possibility that the outer 
tails are in fact
some collimated flows emanating from the pulsar 
magnetosphere, 
such as two jets 
aligned with the pulsar's spin axis near the pulsar
and bent by the 
ram pressure at larger distances.
This interpretation implies a large angle $\Theta$ 
between the spin axis and 
pulsar's velocity ($\Theta\gtrsim 60^\circ$--$70^\circ$, as 
follows from Figs.\ 2 and 3)  and a sufficiently large angle $\zeta$
between the spin axis and the line of sight\footnote{ The large value of
$\zeta$ supports the outer gap interpretation of Geminga's
$\gamma$-ray emission 
(Romani \& Watters 2010).}.

{\sl Chandra} observations have shown that jets emanating along the spin 
axes are ubiquitous among
young PWNe (see, e.g., Weisskopf et al.\ 2000; Pavlov et al.\ 2003),
and the spin axis is often 
approximately aligned with the pulsar velocity direction (Ng \& Romani 2007).
The mechanisms of jet formation and collimation are currently not certain.
In the scenario discussed by Benford (1984),
a fraction of electrons created in 
the vacuum gaps above the
magnetic poles and accelerated along the open magnetic field lines is
deflected toward the spin axis and forms a beam collimated by its own
toroidal magnetic field. 
Another mechanism of axial outflow formation has been
discussed by Komissarov \& Lyubarsky (2004),
who assume that the outflow is originally equatorial
and show that the magnetic hoop stress can stop
the outflow in the surface layers of the equatorial disk and redirect 
it into magnetically confined
polar jets. 

If the pulsar were not moving
with respect to the ambient medium, the jet matter would keep flowing
along the spin axis until the jet is destroyed by the interaction with
the medium. 
The ram pressure exerted onto the jets 
of a moving pulsar can bend the jets in the direction opposite to
that of the pulsar's motion, so that the jets are seen as two
tails behind the moving pulsar\footnote{A similar model has been discussed
by Heinz et al.\ (2008) for microquasars.}.

This scenario allows one to explain the observed asymmetry of the Geminga's
outer tails
(see \S2.2.2).
The asymmetry can be associated with the large (but 
different from $90^\circ$) angle $\Theta$ between the spin axis and the 
pulsar's velocity. At such an orientation the angles between the ram pressure
direction and the jet matter velocity directions are acute and obtuse for
the southeastern and northwestern jets, respectively, which means that bending the northwestern jet is more
difficult. An additional reason for the asymmetry
might be a deviation of the pulsar's magnetic field geometry
 from an ideal centered dipole, which
would lead to different structures of the magnetic field at the two poles and
different properties of the two jets.
 Different brightness of the jets, especially in the pulsar vicinity,
 might be caused by Doppler boosting (if
the angle $\zeta$ between the spin axis and line of sight is different from
$90^\circ$), but it is hard to estimate the
Doppler factor and to infer the angles with the current noisy data.

As the bent polar outflow interpretation of the
outer tails requires a large
value of the angle $\Theta$, while $\Theta$ is apparently
small for most pulsars, this interpretation 
 implies that the outer tails are a rare phenomenon,
in agreement with PWN observations 
that have not shown such tails in other PWNe.
It, however, remains to be understood why $\Theta$ is so different for Geminga.
A theoretical study of the expected distribution of this angle using the
physics of the neutron star birth is required to confirm this
explanation.

\subsection{The nature of the axial tail}

The straight, patchy A-tail is seen up to $50''$ ($0.06 d_{250}$ pc)
from the pulsar. 
Its surface
brightness is not only nonuniform but also variable, as we see
from the comparison of the 2007 and 2004 data.
Assuming a nearly isotropic emission
(which, rigorously speaking, implies a nonrelativistic flow), the A-tail
luminosity, $L_X\sim
(1$--$2)\times 10^{29} d_{250}^2$ erg s$^{-1}$
in the 0.3--8 keV band,
is 
$\sim (3$--$6)\times 10^{-6} d_{250}^2$ 
of  Geminga's spin-down power $\dot{E}$.
The A-tail luminosity
 is $\sim 0.03$--0.06 of the nonthermal (magnetospheric)
 luminosity of the
Geminga pulsar in the same energy range and is
a factor of 2--4 lower than the total luminosity of the outer tails.

There are three 
conceivable explanations of the A-tail:
a jet emanating from the pulsar magnetosphere in
the direction opposite to the pulsar velocity, 
a tail part of the bowshock-tail PWN created by the supesonic motion of the pulsar,
and a Doppler-boosted image of 
a shell into which a fraction of the relativistic PW is directed.
We will discuss these interpretations below, taking into account their
connection with the above-discussed interpretations of the outer tails.

\subsubsection{The axial tail is a pulsar jet?}

The interpretation of the A-tail as a pulsar jet, suggested by P+06, 
is consistent with only one of the above-discussed interpretations of
the outer tails, namely, the hypothesis that
the outer tails represent a boundary of a shell (e.g., an equatorial outflow
bent by the ram pressure). 
As pulsar jets emerge 
along the pulsar's spin axis,
and the A-tail is aligned with the pulsar's trajectory in the sky,
the jet interpretation of the A-tail implies that the spin
axis is likely aligned with the pulsar's velocity. This suggests
that the ``natal kick'' of the Geminga pulsar was directed along
the spin axis,
which has 
important implications for the mechanisms of supernova explosion and 
neutron star formation (e.g., Ng \& Romani 2007).

The lack of a clear (counter)jet ahead of the pulsar (see, however, the note
at the end of \S\,2.2.3) could be explained by Doppler boosting
(the approaching jet is brighter than the receding counterjet, assuming
the jet material flows with nearly relativistic velocities). Alternatively,
the counterjet can be partially or fully destroyed
by the ISM ram pressure, or the outflows in the opposite
directions may be intrinsically
different. 

As described in \S\,2.2.1, the surface brightness is distributed
nonuniformly along the A-tail, with some ``blobs'' seen at different
positions in the images of 2004 and 2007, and a ``gap'' between the
pulsar and the beginning of the A-tail. This means that there are
regions of the enhanced magnetic field and/or higher density along the A-tail,
which might be
 caused by
discrete ejections from the magnetosphere, or they could be
 manifestations of some instabilities (e.g., the sausage instability, 
as discussed by Pavlov et al.\ 2003 for the outer jet of the Vela PWN) 
or internal shocks in the jet flow.
Particularly interesting is
the brightest blob C seen at the apparent end of the A-tail
in the image of 2007 (see Fig.\ 2).
Although a background K star could contribute to the blob C emission
(see \S\,2.2.1), our analysis suggests that the star's contribution is
not dominant and blob C could be associated  with the jet's 
termination shock.   To understand the nature
of the blobs, it would be important to study their evolution
in a series
of deep observations,  which, in particular, would help 
estimate the flow speed in the jet.

Using the observed diameter of the alleged jet and the estimate for the
energy injection rate,
 P+06 estimate
the jet's magnetic field: $B_{\rm jet}
\sim 100\, 
\mu{\rm G}$. For such a magnetic field the expected jet length is
$l_{\rm jet}\sim v_{\rm jet}\tau_{\rm syn}\sim 0.6\, (v_{\rm jet}/0.5c)\,(B_{\rm jet}/100\,\mu{\rm G})^{-3/2}$ pc, where $v_{\rm jet}$ is the bulk flow velocity in
the jet, and $\tau_{\rm syn}$ is the synchrotron cooling time.
The jet length 
estimated
from the initial bright portion of the axial tail, 
$l_{\rm jet}\sim 0.06\, d_{250}/\sin i$ pc,  is much smaller than this
value unless $v_{\rm jet}$ is much smaller than $0.5 c$ and/or $\sin i$ is
small, which seems unlikely. To explain this contradiction, P+06 speculate
that 
the jet becomes uncollimated or 
destroyed well before it radiates its entire internal energy. 
To check such speculations and test the pulsar jet interpretation,
deeper observations are required.

\subsubsection{
The axial tail is a shocked pulsar wind?}

A tail-like structure similar to the observed A-tail could form
behind the supersonically moving pulsar due to the collimation of the
shocked PW by the ram pressure (B+05; Romanova et al.\ 2005). 
For instance,
if we assume that 
 the outer tails and the possible
arc ahead of the pulsar delineate the CD surface (see \S\,3.1.1),
then the A-tail might be the shocked PW
 immediately outside (and perhaps behind) the bullet-like TS.
P+06 have 
shown that this interpretation 
is not quantitatively consistent with the available simulations of 
bowshock-tail PWNe, but those simulations do not take into account the
intrinsic anisotropy of the PW.
In addition, it would be difficult to explain the presence of the blobs
and the variability of the A-tail in the framework of this interpretation.
Therefore, we consider this interpretation unlikely.

If the outer tails are bent polar outflows (\S\,3.1.2), then
the ram-pressure confined shocked PW would be the only
possible explanation for the A-tail. A more detailed interpretation
of the A-tail would depend on the PW model. For instance, if the
PW were intrinsically isotropic, then the A-tail might
be interpreted as originating
from the shocked PW ``sheath" immediately outside the TS,
and the observed
width of the tail would imply the distances $R_{\rm TS,h} \approx 3''$
and $R_{\rm CD,h} \approx 4''$
of the TS and CD heads from the pulsar. Therefore,
we would expect a bright arc (brighter than any part of the axial tail) 
$\approx (3''$--$4'') \sin i$ ahead of the pulsar.
No such a bright arc is seen in the images, but this
 does not necessarily rule out
the TS origin of the A-tail because the TS head could be closer to
the pulsar and hidden within the pulsar image if the PW is anisotropic
(P+06).
In this interpretation, however, we have to assume that no emission
is seen from the CD surface region, which looks somewhat unnatural.

In the bent polar outflow interpretation of the outer tails,
one could also assume that 
the A-tail tail is associated with the
CD-confined cylindrical region behind the (unresolved) TS.
In this case, for an isotropic PW,
we would expect a bright arc 
$\approx 1''\sin i$ ahead of the pulsar,
which can easily be hidden within the pulsar image.
In the framework of this interpretation, the non-uniformity and 
variability of the A-tail could be explained by shear instabilities
at the CD surface (cf.\ \S\,3.1.1), which could also decelerate the
flow,  so that the blobs' velocity would be lower than that in the jet
interpretation of the A-tail.
We should note, however, that in such interpretations the polar outflows are
more luminous than the ram-pressure confined tail (perhaps an equatorial
outflow), which has not been observed
for any other pulsar. 

\subsubsection{
The axial tail is a Doppler-boosted shell projection?}

As we have mentioned above,
the image of
 a shell formed by material
outflowing with relativistic speeds may be strongly affected by
Doppler boosting,  which brightens those parts
of the shell where the angle $\theta$ between the bulk flow velocity and the
line of sight is the smallest.
For instance,
a conical shell, in which the material flows from the
cone vertex at the pulsar position,
would look like a straight strip\footnote{ 
A similar model has been discussed by Radhakrishnan \& Deshpande (2001) 
who interpret the ``inner jets'' of the Vela PWN as a projection of the
rotating beams of relativistic particles emitted along the magnetic axes.}
corresponding to the minimum $\theta$.
 If the
material flows from the head of a paraboloid-like shell (e.g., the
CD surface), the observer would see a shorter strip, 
detached from the paraboloid head in general case. 
In principle, one could imagine that
the A-tail is such a projection of the shell formed by the
shocked PW that flows out with relativistic velocities between the
TS and CD, while the outer tails are bent polar outflows (jets).
In this interpretation, the true transverse radius of the shell would be
larger than the observed width of the A-tail. 
The blobs in the A-tail might be some local instabilities
in the relativistic flow, which would likely move with relativistic
bulk-flow velocities. Therefore, it would be important to examine the
blob motion in future observations.

\section{Summary}

The new {\sl Chandra} and {\sl XMM-Newton} observations of the Geminga PWN
have confirmed that it has three tail-like components, 
unlike any other detected PWN.
The new observations have allowed us to image the tails at larger distances from
the pulsar and establish their patchy structure. Comparing the new and previous
 {\sl Chandra} observations, we have found indications of PWN  variability,
especially in the axial tail and the emission ahead of the pulsar.
 In particular, we found up to three
blobs in the axial tail, at different positions in 2004 and 2007.

Similar to other X-ray PWNe, the Geminga PWN 
is due to
 synchrotron radiation of shocked PW comprised of relativistic particles.
Based on the new 
and old observations,
we have proposed several competing interpretations of the PWN structure.
Very likely, the outer tails delineate
a limb-brightened
boundary of a shell-like region of interaction of the shocked PW and shocked
ISM, while the axial tail is a  pulsar
jet along the spin axis aligned with the pulsar's trajectory.
Such an interpretation implies a nonrelativistic speed of the bulk outflow
along the shell, possibly decelerated by the shear instability and mass loading.
Alternatively, 
the outer tailis could be
polar outflows from the pulsar magnetosphere (e.g., 
pulsar jets along the spin axis), bent by the ISM ram pressure,
in which case the axial tail could be a shocked PW (e.g., an equatorial
outflow) collimated by the ISM ram pressure exerted on the supersonically
moving PWN. 

To discriminate between various interpretation of the observed PWN,
a series of carefully designed
 \chan\ observations is required. In particular, such observations should 
allow one
%, in particular, to
to measure the speeds of the bulk flows in the tails,  which would distinguish
fast jets from 
ram-pressure-confined pulsar winds slowed down by the 
interaction with the ambient ISM.
Also, such observations should be deep enough
to establish the true morphology 
%and 
%variability 
of the emission in the immediate vicinity of the pulsar.
For instance, if a deeper observation convincingly shows that there is an arc ahead
of the pulsar connecting the two outer tails, then the bending axial outflows
scenario will be ruled out. If, however, we see 
%that the outer tails
%originate from the pulsar, the interpretation would depend on the
%tails' shape in the immediate vicinity of the pulsar. For instance, if the
two straight tails originating from the pulsar in a direction 
%is clearly not
%perpendicular 
inclined to the pulsar's velocity direction, then the tails can be
interpreted as bent jets.
In addition, the detailed modeling of anisotropic
magnetic PW from a high-speed pulsar will also be 
extremely useful to properly interpret the observational data.

\acknowledgments 
We thank Andrew Melatos for useful discussions of the PWN modeling.
Support for this work was provided by the National Aeronautics and Space 
Administration through {\sl Chandra} Award Number GO7-8053A issued 
by the {\sl Chandra} X-ray Observatory Center, which is operated by the 
Smithsonian Astrophysical Observatory for and on behalf of the National 
Aeronautics Space Administration under contract NAS8-03060. The work by
GGP was also partially supported by NASA grant NNX09AC84G. 

{}

\clearpage

{\hoffset=-1cm
\begin{deluxetable}{lrccccc}
\tablecolumns{7} \tablewidth{0pc} \tablecaption{Properties 
of the 
PWN elements from the {\sl Chandra} data of 2007 and 2004.}
\tablehead{PWN element & Area\tablenotemark{a} & Counts\tablenotemark{b}
& $S/N$
& $\Gamma$
& Flux\tablenotemark{c} & Luminosity\tablenotemark{d} 
}
\startdata
A-tail (2007; bright)\tablenotemark{e}
& 705.6 & $82.6\pm12.0$ & $6.9$ & $1.84^{+0.29}_{-0.26}$
        & $1.19^{+0.30}_{-0.32}$ & $0.92^{+0.15}_{-0.12}$ \\
A-tail (2007; bright, w/o blob C)\tablenotemark{f}
& 
677.1  & $56.8\pm10.8$  & 5.3 & $1.87^{+0.41}_{-0.36}$ 
	& $0.77^{+0.20}_{-0.21}$ & $0.60^{+0.14}_{-0.10}$  \\
A-tail (2007; faint)
& 2404.2 & $29.8\pm17.2$ & 1.7 & $-0.12^{+0.86}_{-1.18}$
        & $1.15^{+0.35}_{-1.15}$ & $0.86^{+0.53}_{-0.46}$ \\
A-tail (2004) & 118.1 & $45.6\pm7.1$ & 6.4 & $1.26^{+0.28}_{-0.27}$
        & $2.15^{+0.66}_{-0.78}$ & $1.64^{+0.35}_{-0.31}$  \\
N-tail (2007; bright)
& 462.3 & $39.4\pm8.8$ & 4.5 & $-0.11^{+0.49}_{-0.58}$  
	& $1.21^{+0.25}_{-0.41}$ & $0.91^{+0.32}_{-0.24}$  \\
N-tail  (2007; faint)
 & 4011.9 & $52.9\pm24.3$ &  2.2 & $0.82_{-0.57}^{+0.60}$ & $1.30_{-1.07}^{+0.64}$ & $0.98_{-0.43}^{+0.50}$ \\
N-tail  (2007; entire)
& 4474.2 & $92.3\pm26.9$ & 3.4 & $0.47^{+0.49}_{-0.53}$  
	& $2.33^{+0.63}_{-1.45}$ & $1.76^{+0.64}_{-0.63}$ \\
S-tail  (2007; bright)
& 2258.0 & $139.0\pm19.6$ & 7.1 & $1.17^{+0.27}_{-0.25}$  
	& $2.58^{+0.42}_{-0.50}$ & $1.96^{+0.38}_{-0.34}$  \\
S-tail (2007; faint)
 & 3740.7 &  $28.8\pm22.7$ &  1.3 & $1.82_{-0.46}^{+0.50}$ & $1.03_{-0.36}^{+0.48}$ &    $0.80_{-0.22}^{+0.26}$ \\
S-tail (2007; entire)
& 5998.7 & $167.8\pm33.8$ & 5.0 & $1.59^{+0.30}_{-0.28}$  
	& $3.21^{+0.72}_{-0.70}$ & $2.47^{+0.46}_{-0.38}$  \\
S-tail (2004) & 985.3 & $33.0\pm9.0$ & 3.7 & $1.20^{+0.45}_{-0.42}$
        & $1.58^{+0.53}_{-0.57}$ & $1.20^{+0.49}_{-0.38}$ \\
Arc (2007) & 86.6 & $16.3\pm4.8$ & 3.4 & $1.07^{+0.73}_{-0.67}$ 
	& $0.25^{+0.19}_{-0.06}$ & $0.19^{+0.10}_{-0.07}$ \\
Arc (2004) & 86.6 & $22.8\pm 5.1$ & 4.5 & $0.90^{+0.46}_{-0.43}$ 
	& $1.19^{+0.51}_{-0.83}$ & $0.90^{+0.39}_{-0.28}$ \\
\enddata
\tablecomments{The initial bright and extended
faint portions of the three PWN tails in the data
of 2007 are shown in Fig.\ 3. The A-tail, N-tail, and S-tail stand for
the axial tail, the northern tail, and the southern tail, respectively.
The spectra of the PWN elements were fitted with the absorbed power-law model
(photon index $\Gamma$) at a fixed hydrogen column density 
$N_{\rm H}=1.1\times 10^{20}$ cm$^{-2}$. The errors are given
at the $1\sigma$ level, for one interesting parameter.
}
\tablenotetext{a}{Area of the PWN element in arcsec$^2$.}
\tablenotetext{b}{Background-subtracted counts in PWN elements.}
\tablenotetext{c}{Observed (absorbed) flux in the 0.3--8 keV band,
 in units of $10^{-14}$ erg cm$^{-2}$ s$^{-1}$, estimated with XSPEC. }
\tablenotetext{d}{Unabsorbed luminosity in the 0.3--8 keV band, assuming isotropic
emission and distance of 250 pc, in units of $10^{29}$ erg s$^{-1}$.}
\tablenotetext{e}{Blob C at the end of the brighter portion of the axial tail is included.}
\tablenotetext{f}{
Blob C ($3''$ radius) is excluded from the data.}
\end{deluxetable}
}

\clearpage
\begin{figure}
\centering
\includegraphics[width=6.8cm,angle=0]{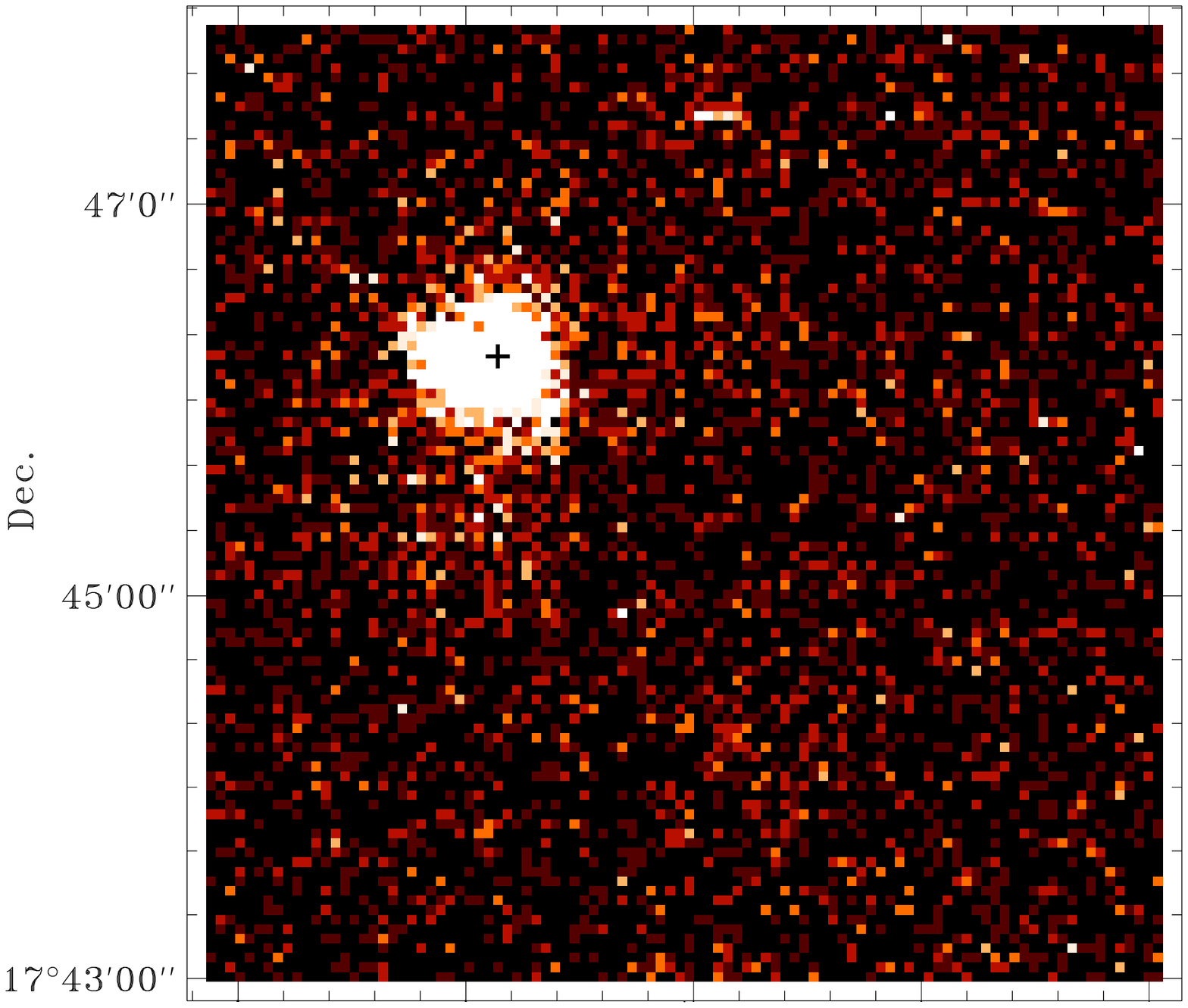}
\includegraphics[width=6.8cm,angle=0]{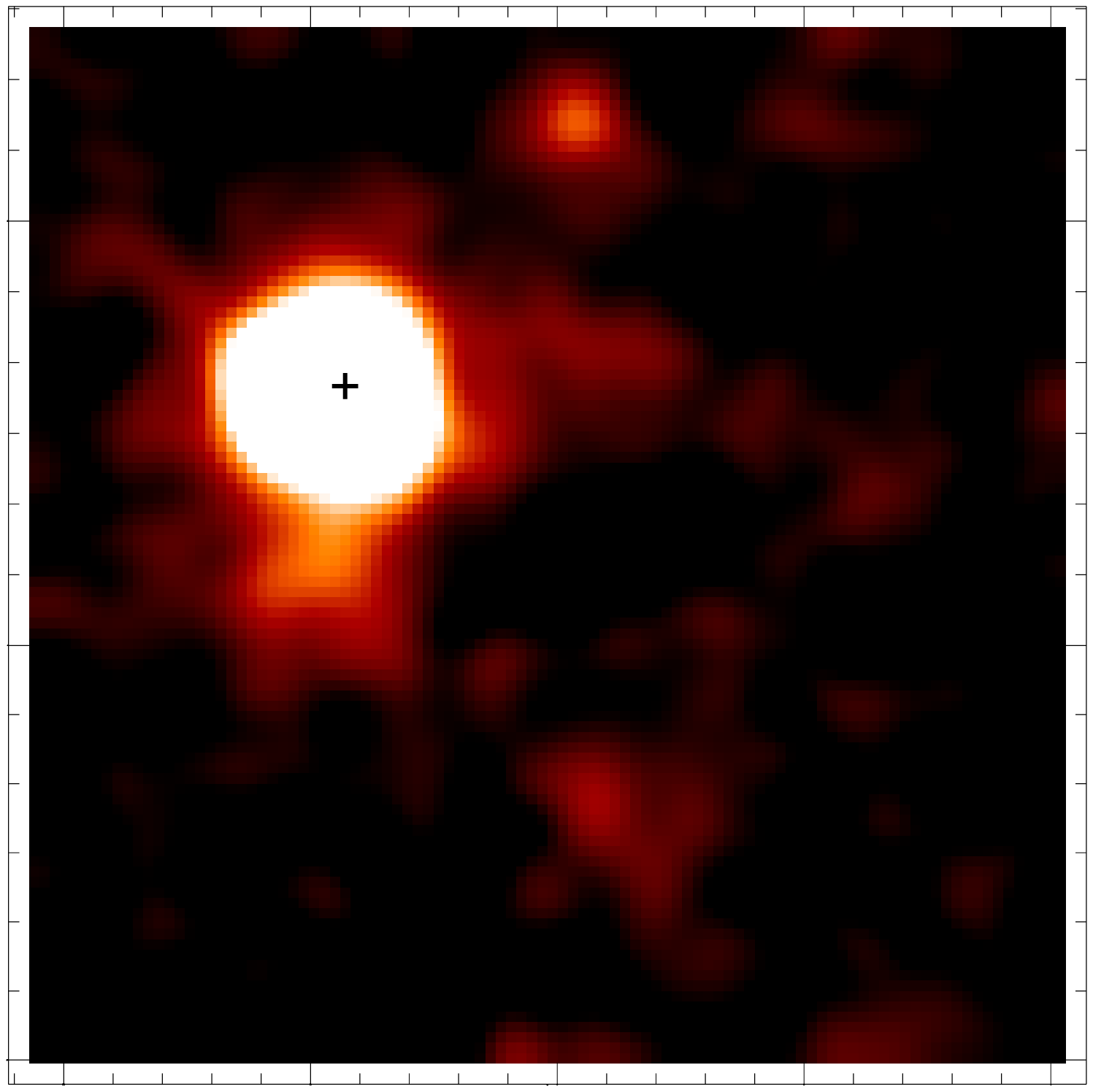}\\
\includegraphics[width=6.8cm,angle=0]{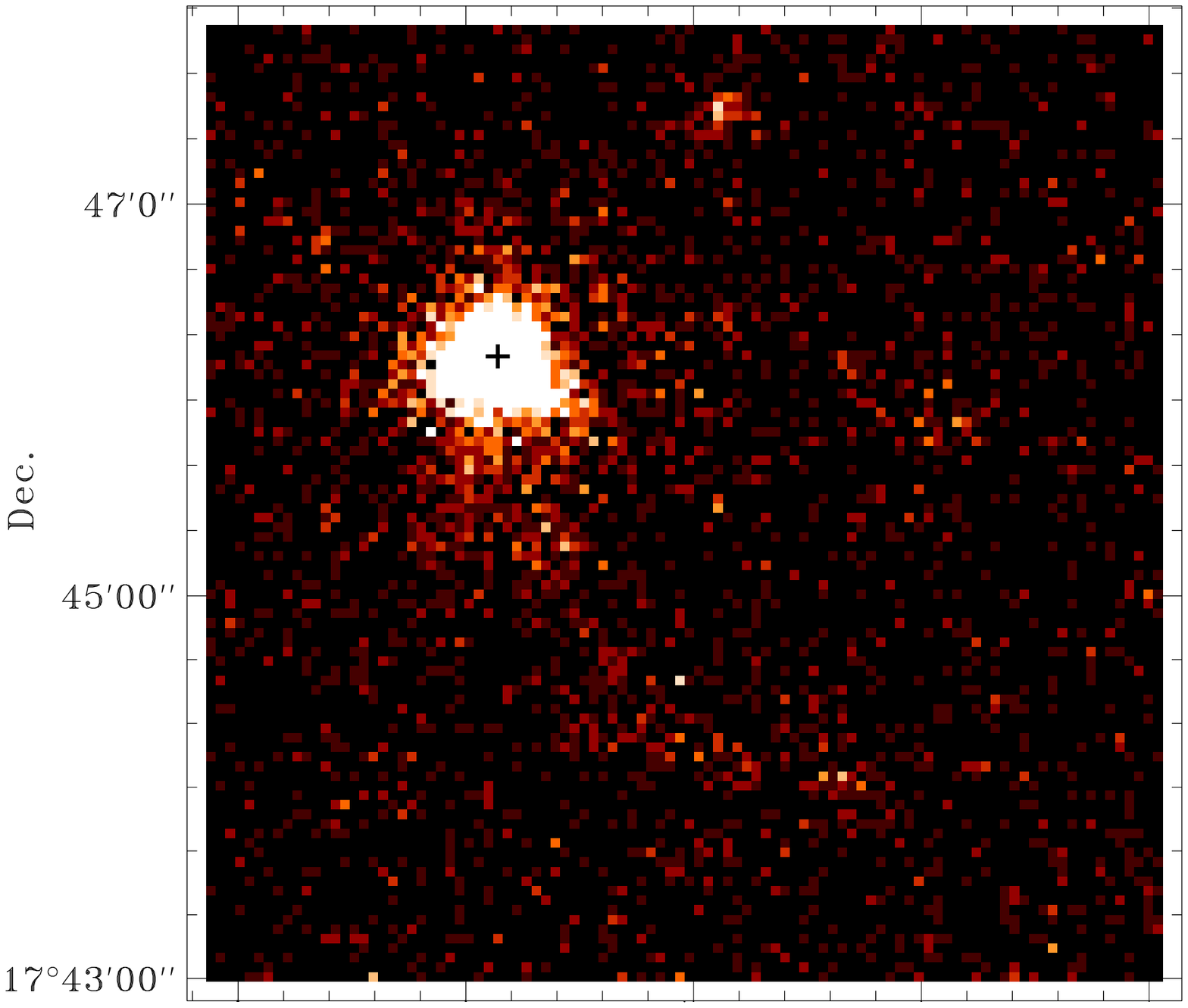}
\includegraphics[width=6.8cm,angle=0]{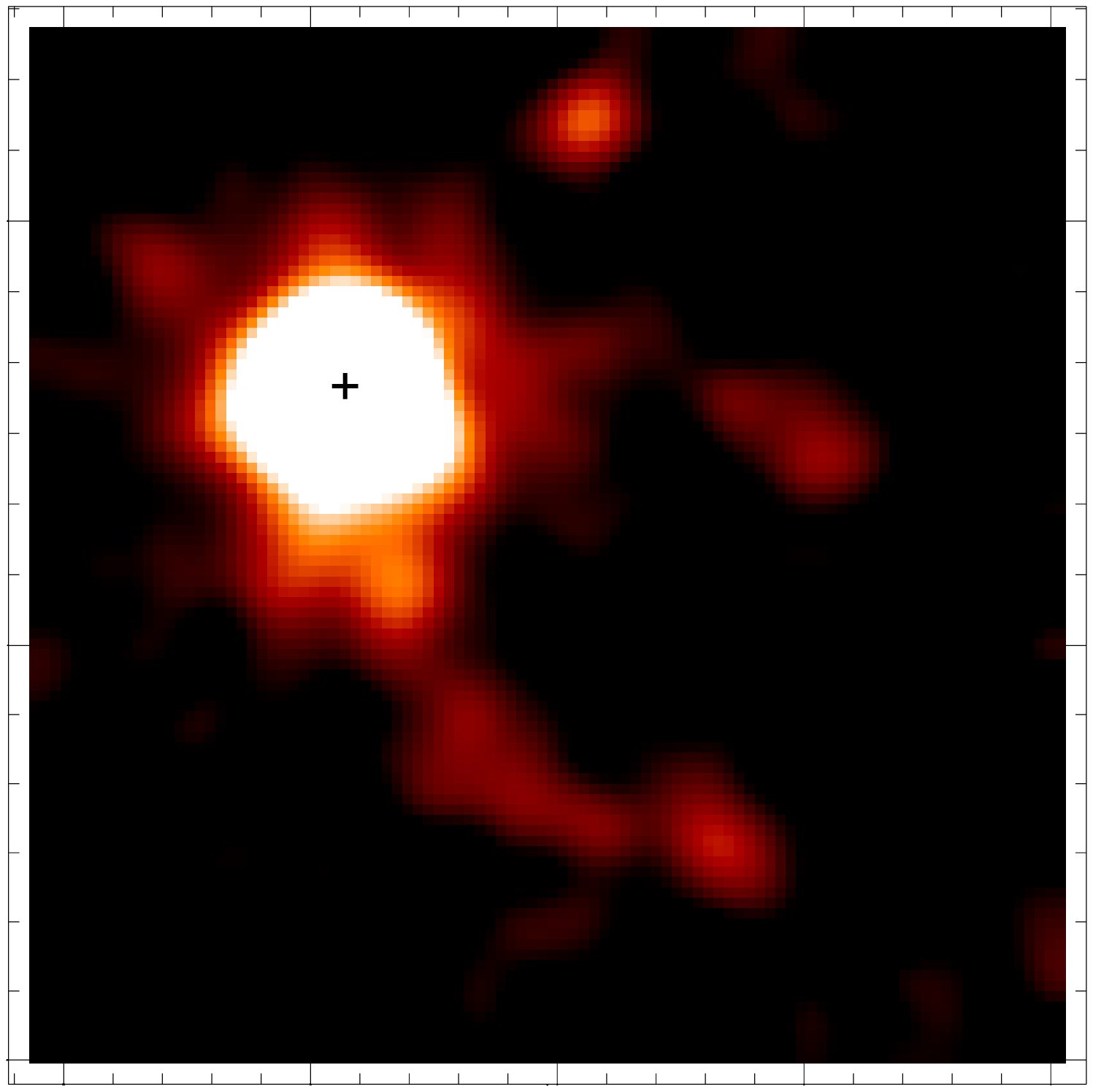}\\
\includegraphics[width=6.8cm,angle=0]{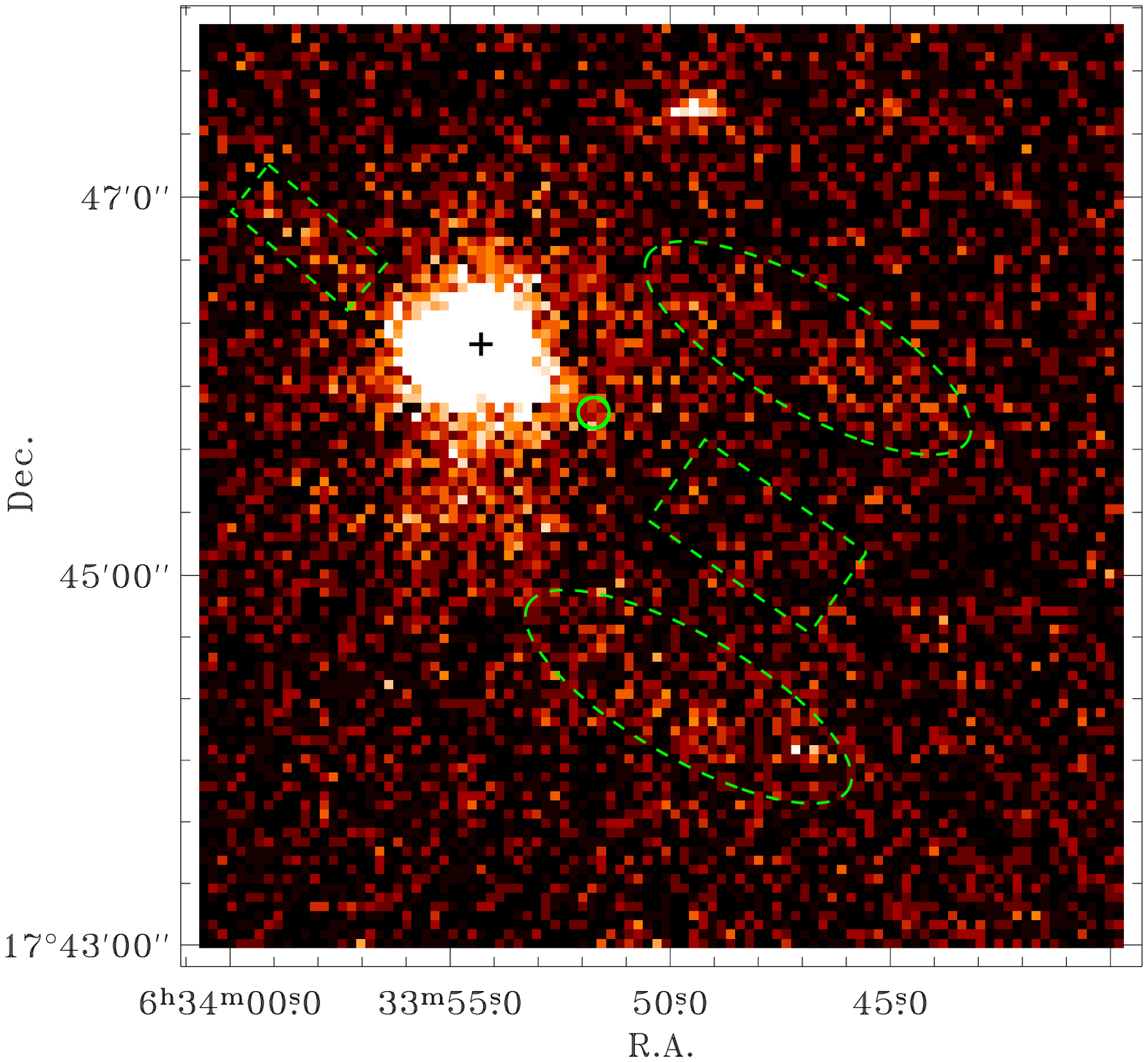}
\includegraphics[width=6.8cm,angle=0]{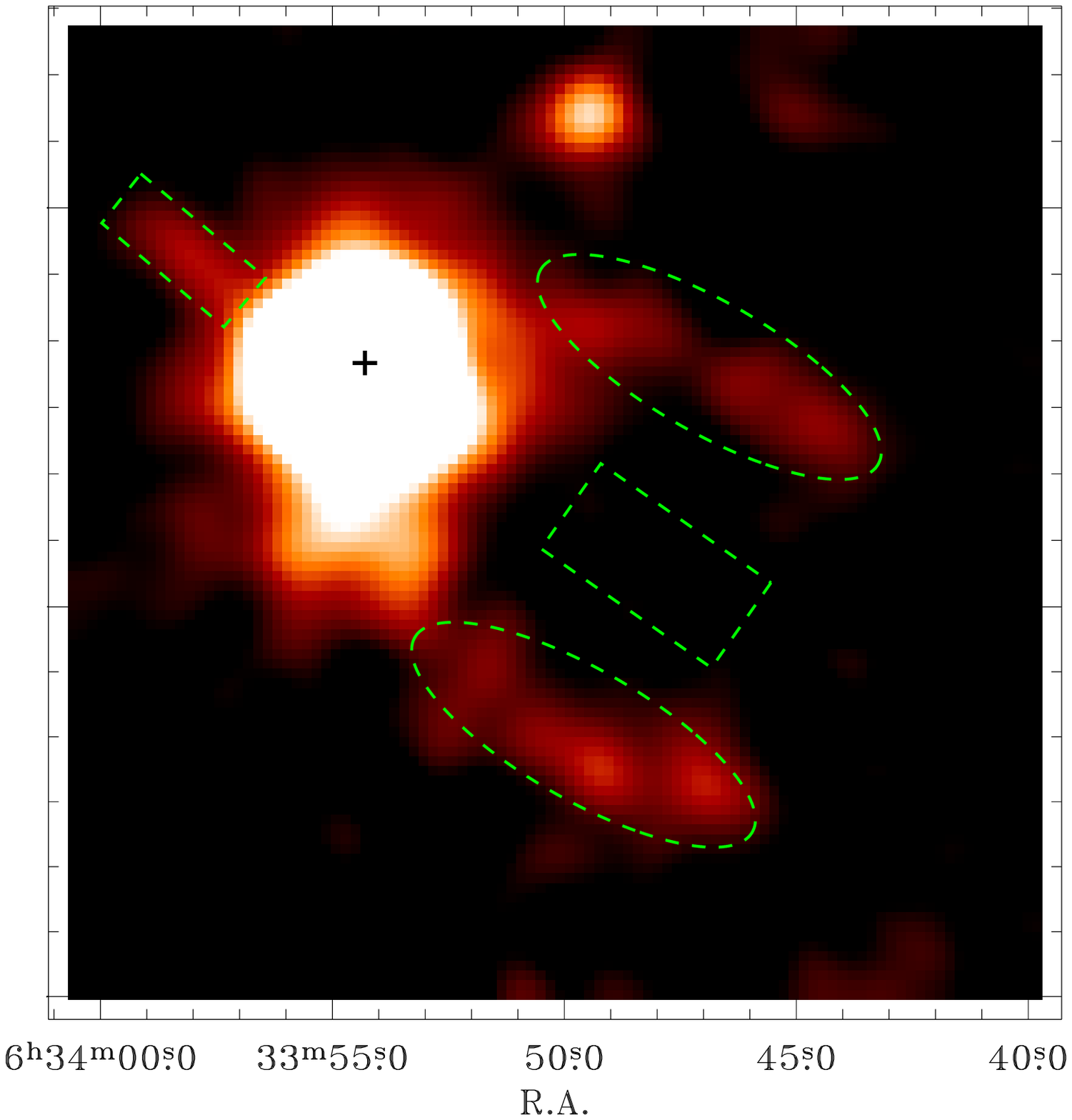}
\vspace{0.8cm}
\caption{{\sl XMM-Newton} MOS1+MOS2 images ($5'\times 5'$) 
of the Geminga pulsar and 
its PWN in the 0.5--8 keV band. 
The top, middle and bottom panels correspond to the observation of
2004--2007 
(64 ks total scientific exposure), 
2002 (78 ks), and 2002--2007 (142 ks), respectively.
The images in the left panels are binned in 
$3''\times 3''$ pixels, while the images in the right panels are
additionally 
smoothed with a $18''$ FWHM Gaussian. 
The ellipses ($120''\times 40''$) show the regions for which the
spectra and fluxes were measured, while the $64''\times 32''$ rectangle
between the ellipses 
was used for estimating the upper limit on the surface brightness
between the outer tails (see \S\,2.2.2).
The $50''\times 20''$ rectangle ahead of the pulsar was used to measure
the spectrum of the ``streak'' (see \S\,2.2.3).
The $5''$ radius circle in the bottom left panel is centered at the position
of blob C found in the \chan\ observation of 2007 (see Fig.\ 2 and \S\,2.2.1).
The source north-northwest of the pulsar is an 11-th magnitude K star (C+03).
}
\end{figure}

\clearpage
\begin{figure}
\centering
\includegraphics[height=7.8cm,angle=0]{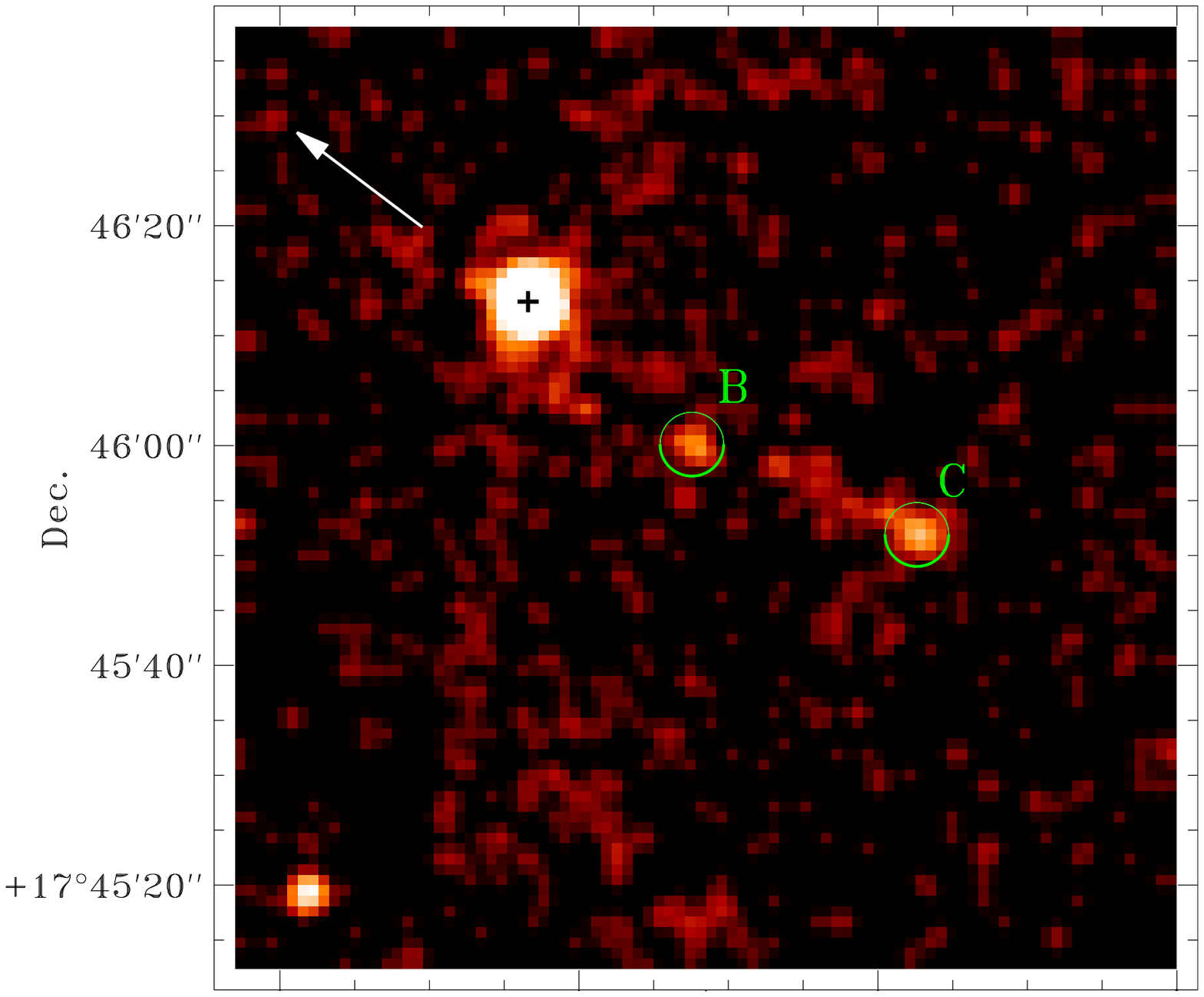}\\
\includegraphics[height=7.8cm,angle=0]{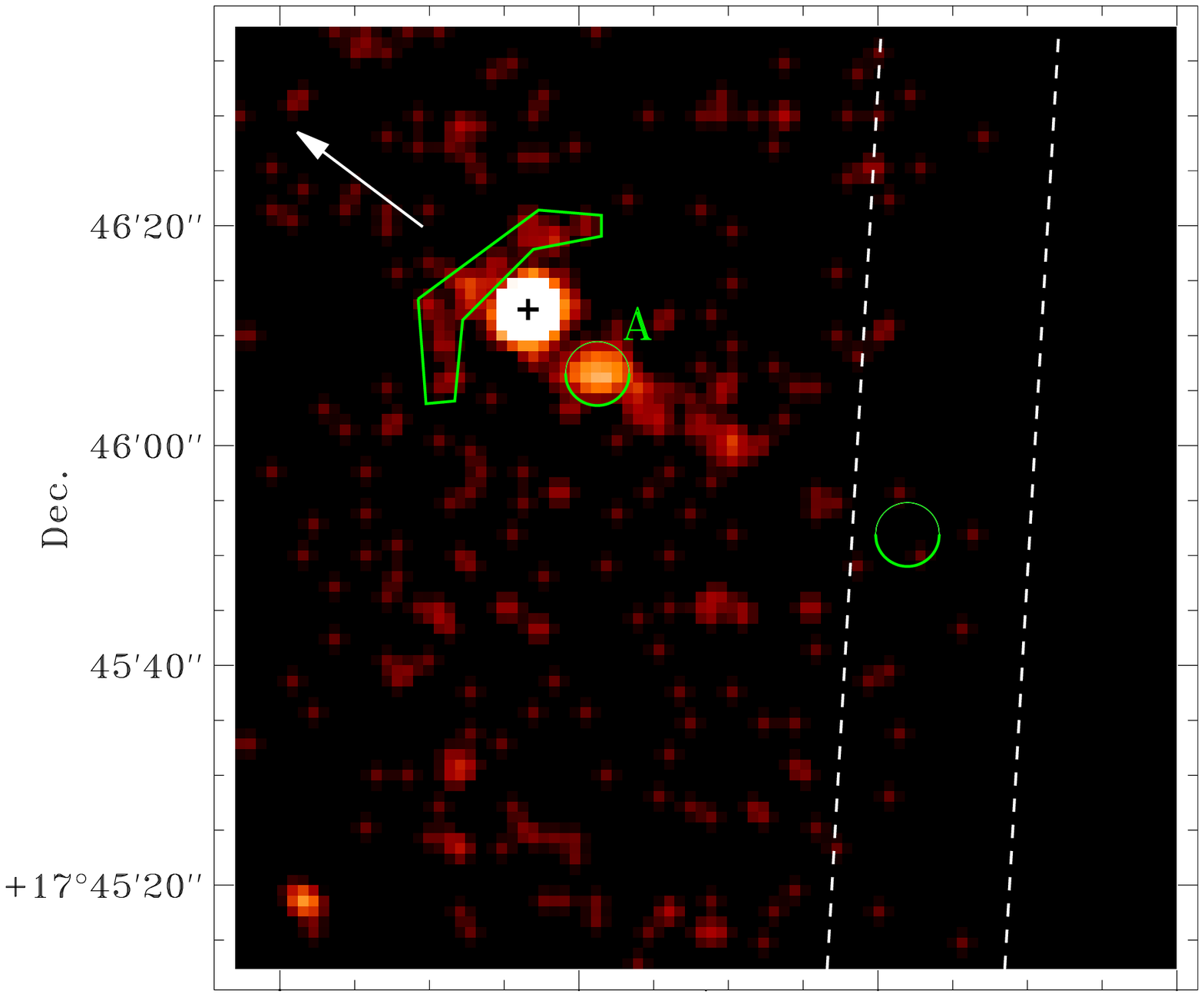}\\
\includegraphics[height=7.8cm,angle=0]{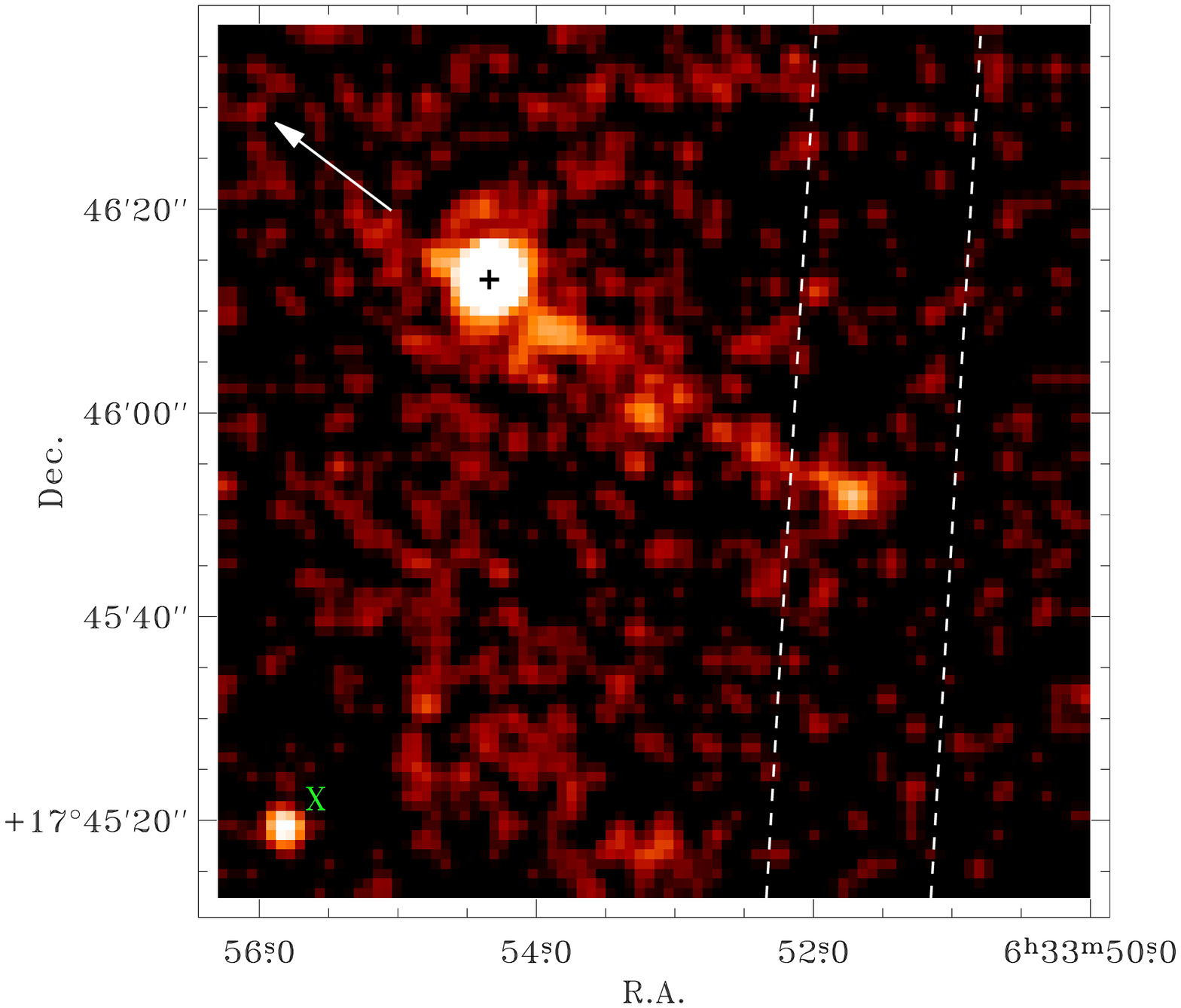}
\vspace{0.8cm}
\caption {
$1\farcm 6\times 1\farcm 6$ images of the Geminga pulsar and its PWN from the {\sl Chandra} observations
of 2007 and 2004 (top and middle
panels, respectively), and the combined image (bottom), in the 0.3--8 keV range. 
The original data were binned in $1''\times 1''$ pixels and smoothed 
with a $2''$ FWHM Gaussian. The arrows show the pulsar's proper motion,
whereas the dashed lines in the middle and bottom panels 
indicate the sky region
for which the exposure 
was reduced by the telescope dithering (from 18.8 to 0 ks, in the 2004 observation).
Three $3''$-radius  circles mark ``blobs'' in the A-tail
(see \S\,2.2.1). 
The label X marks a field star (see P+06). 
}
\end{figure}

\clearpage
\begin{figure}
\centering
\includegraphics[width=4.7in,angle=0]{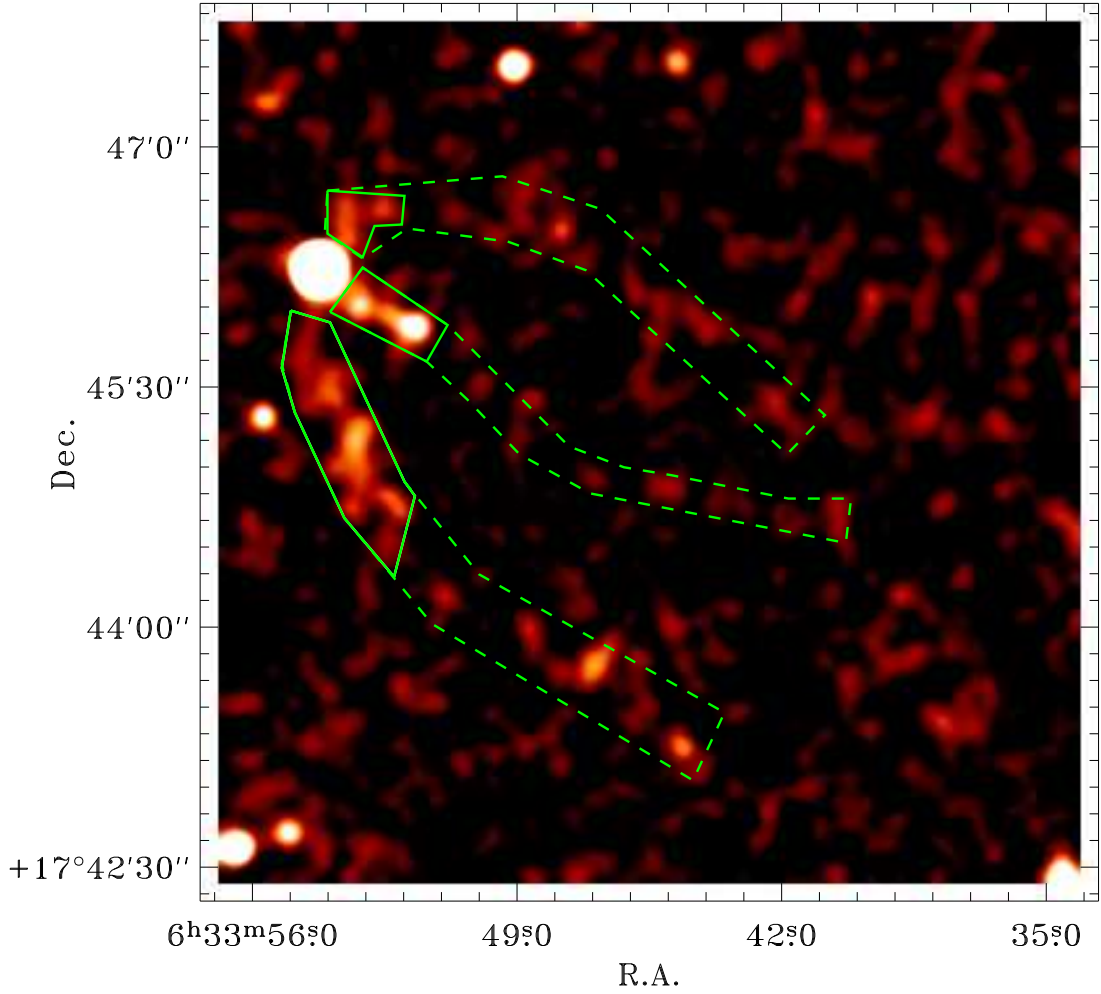}
\vspace{1.2cm}
\caption {
$5.4'\times 5.4'$ image of the Geminga PWN in the \chan\ data of 2007.
The solid and dashed contours show the regions used for extracting 
the spectra of brighter and fainter
parts of the PWN tails, respectively.
}
\end{figure}
%
%\clearpage
\begin{figure}
\centering
\includegraphics[width=4.7in,angle=0]{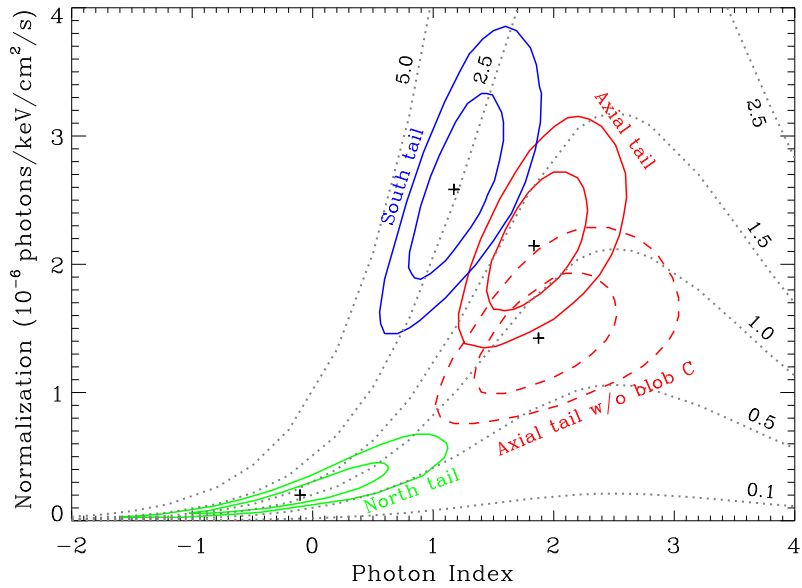}
\vspace{-0.0cm}
\caption {Confidence
contours for the PL normalization versus photon index
(at the 68.3\% and 95.4\% levels,
for two interesting parameters)
for bright portions of the PWN tails from the {\sl Chandra} 2007 data.
The hydrogen column density is fixed at
$N_{\rm H}=1.1\times 10^{20}$ cm$^{-2}$.
The dotted curves
correspond to constant unabsorbed flux values
(depicted near the curves, in units of $10^{-14}$ erg cm$^{-2}$ s$^{-1}$).
}
\end{figure}

\clearpage
\begin{figure}
\centering
\includegraphics[width=5in,angle=0]{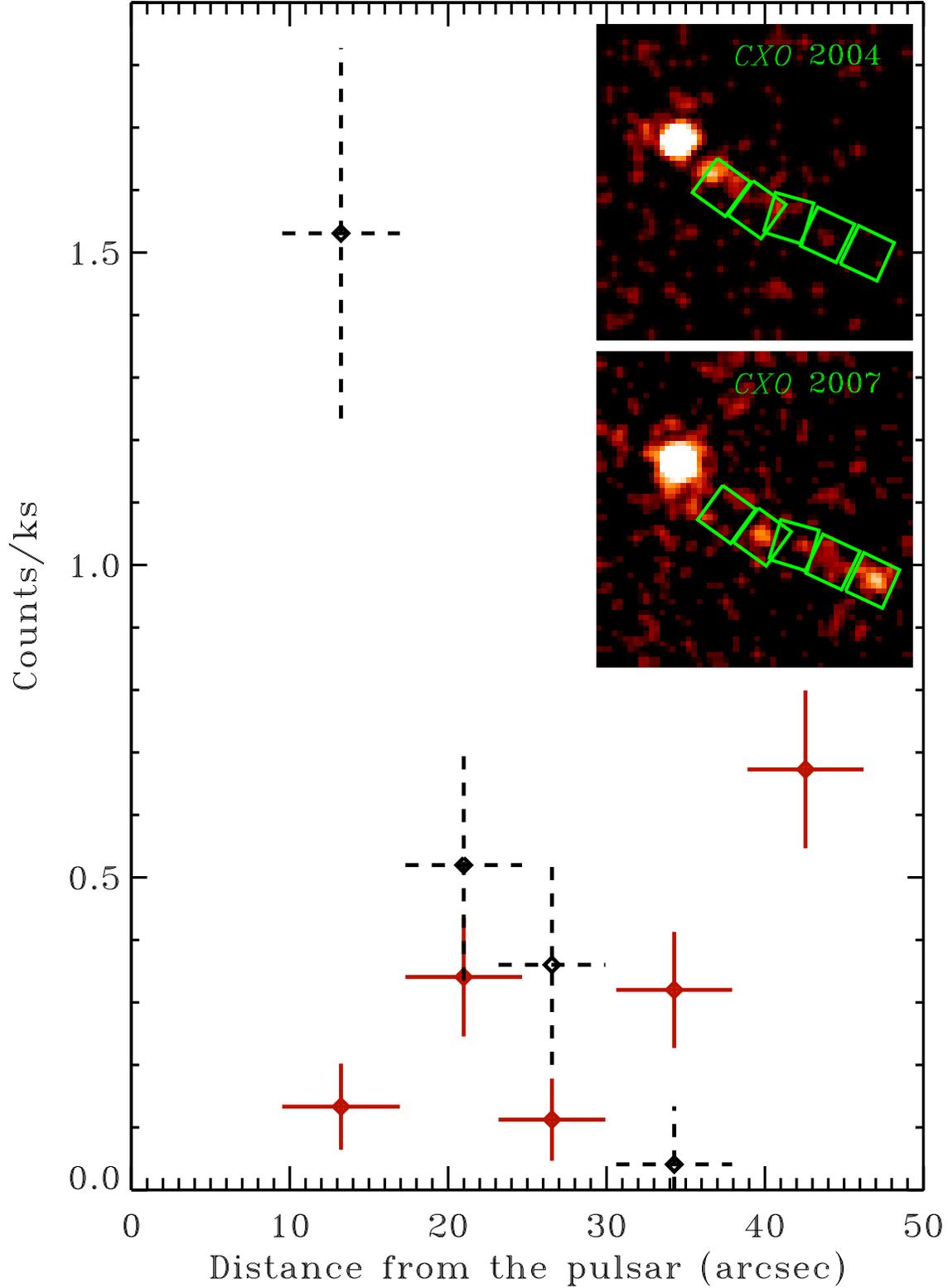}
\vspace{2.0cm}
\caption {Count rate profiles along the axial tail in the {\sl Chandra} data.
The net count rates are calculated in 
square boxes of equal size (59.5
arcsec$^2$), shown in the inset. The dashed (black) and solid (red) 
points (with $1\sigma$ errors) are for the 2004 
and 2007 data, respectively.
The count rates and their errors for the 2007 data are multiplied by a factor of
1.6 to account for the different sensitivities of the I3 and S3 chips.
}
\end{figure}

\clearpage
\begin{figure}
\centering
\includegraphics[width=5in,angle=0]{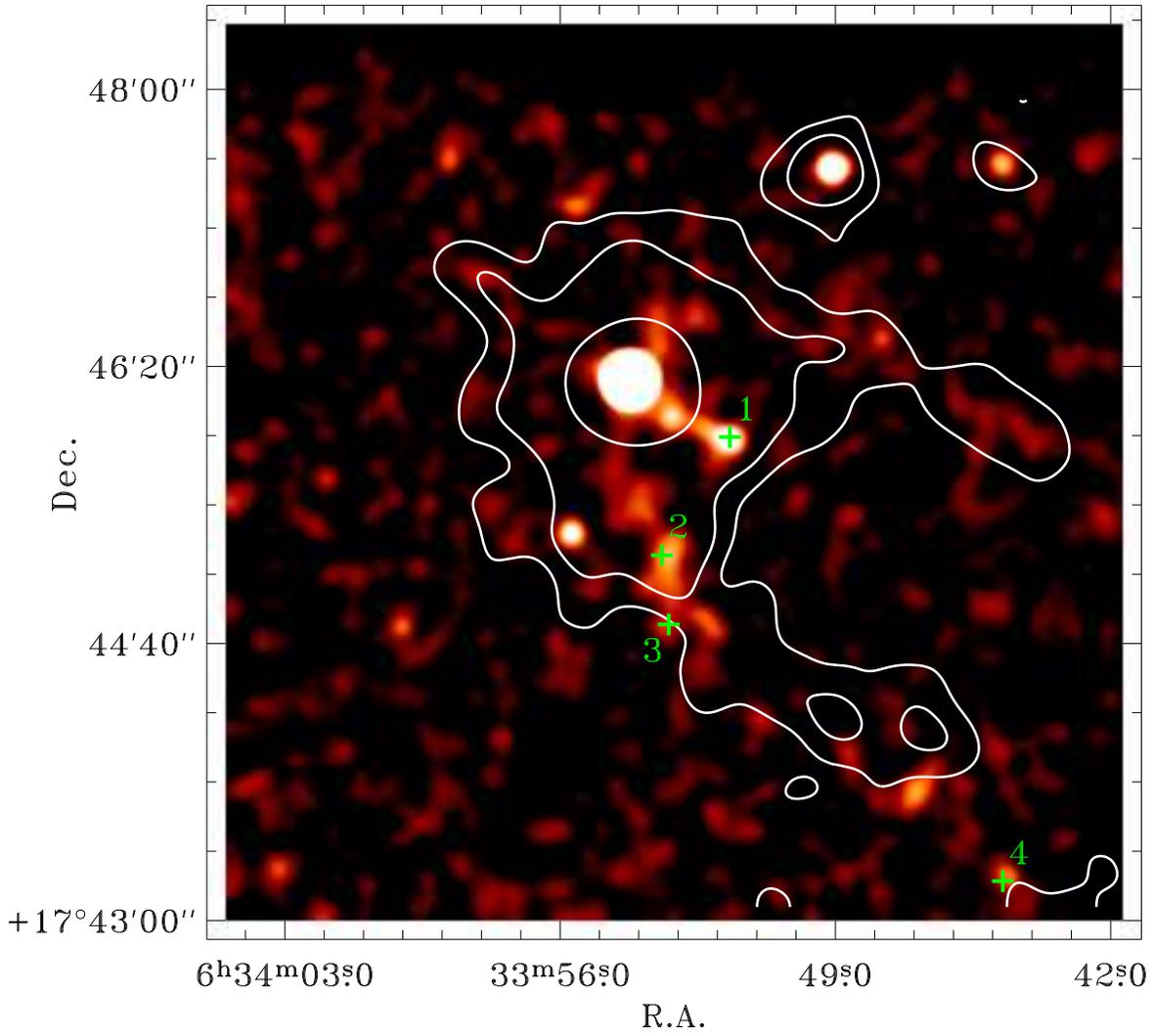}
\vspace{2.0cm}
\caption {
 $5\farcm4\times 5\farcm4$
image of the Geminga pulsar and its PWN
from the {\sl Chandra} data of 2007 binned in $1''\times 1''$ pixels and 
smoothed with an $8''$ FWHM Gaussian,
 with overlaid brightness
contours from the {\sl XMM-Newton} image shown in Fig.\ 1.
The green crosses with numbers indicate 
the positions of optical stars projected
onto the PWN elements.
}
\end{figure}

\end{document}